# Interpretable Scalar-on-Image Linear Regression Models via the Generalized Dantzig Selector

Sijia Liao
Statistics & Data Science GIDP, University of Arizona

Xiaoxiao Sun [*]
Department of Epidemiology and Biostatistics, University of Arizona

Ning Hao
Department of Mathematics, University of Arizona

Hao Helen Zhang
Department of Mathematics, University of Arizona

June 12, 2026

**Abstract**

The scalar-on-image regression model examines the association between a scalar response and a bivariate function (e.g., images) through the estimation of a bivariate coefficient function. Existing approaches often impose smoothness constraints to control the bias-variance trade-off, and thus prevent overfitting. However, such assumptions can hinder interpretability, especially when only certain regions of an image influence changes in the response. In such a scenario, interpretability can be better captured by imposing sparsity assumptions on the coefficient function. To address this challenge, we propose the Generalized Dantzig Selector, a novel method that jointly enforces sparsity and smoothness on the coefficient function. The proposed approach enhances interpretability by accurately identifying regions with no contribution to the changes of response, while preserving stability in estimation. Extensive simulation studies and real data applications demonstrate that the new method is highly interpretable and achieves notable improvements over existing approaches. Moreover, we rigorously establish non-asymptotic bounds for the estimation error, providing strong theoretical guarantees for the proposed framework.

[*]Correspondence should be addressed to: xiaosun@arizona.edu.

# 1 INTRODUCTION

In recent years, the *scalar-on-image regression model* has received substantial attention within the statistical research community, catalyzing significant advancements in methodology, theoretical development, and practical applications across diverse fields such as healthcare, neuroscience, and environmental science (Guillas and Lai, 2010; Wang et al., 2017; Kang et al., 2018). This framework is designed to explore the relationship between a scalar response and a bivariate predictor in the form of an image over a spatial domain, providing an interpretable and flexible approach to handling complex data with spatial structures.

Formally, the scalar-on-image linear regression model assumes that we observe a set of paired data $\{x_i(t,s), y_i\}$ for $i = 1, 2, \ldots, n$, where each $x_i(t,s)$ is an image-valued predictor, viewed as a bivariate function defined on the spatial domain $\Omega = J_t \times J_s$, with $J_t, J_s \subseteq \mathbb{R}$ representing intervals on the real line. Correspondingly, $y_i$ is a scalar-valued response. The relationship between the predictor and the response is modeled as

$$y_i = \alpha + \iint_{\Omega} x_i(t,s)\beta(t,s)\, dt\, ds + \epsilon_i, \tag{1}$$

where $\alpha$ denotes the intercept, $\beta(t,s)$ is an unknown coefficient function defined over the spatial domain $J_t \times J_s$, and $\epsilon_i$ represents the random error capturing variability unexplained by the model. The function $\beta(t,s)$ plays a central role in identifying which regions within the image are informative to predicting the response, providing valuable insights for association analysis in real applications. Without loss of generality, throughout this paper we assume $\Omega = [0,1] \times [0,1]$.

There are several fundamental challenges in scalar-on-image regression. We illustrate these using the hurricane prediction problem as a motivating example. The formation of hurricane in the North Atlantic basin is closely linked to the sea surface temperature (SST), only in specific geographic regions, as suggested by Bell and Chelliah (2006) and Saunders and Lea (2008). In other words, the influence of SST on hurricane is heterogenous across the ocean, as show in Figure 1. Consequently, for a hurricane prediction model, the coefficient estimates for regions that do not influence hurricane formation should ideally be exactly zero to enhance interpretability. At the same time, for regions where the climate influence is non-negligible, enforcing smoothness remains crucial as climate variables such as temperature typically exhibit continuous spatial variation. This dual requirement (i.e., sparsity for interpretability and smoothness for stability) highlights the need for refined modeling approaches that can simultaneously address both challenges effectively.

Some of existing techniques emphasize smoothness, such as spline-based approaches (He and Shi, 1996; Ramsay et al., 2007; Crambes et al., 2009; Yuan and Cai, 2010; Wood et al., 2016; Wood, 2017) and functional principal component analysis methods (Yao and Lee, 2006; Reiss and Ogden, 2007, 2010; Goldsmith et al., 2011), and they often produce diffuse and less localized coefficient surfaces, potentially obscuring meaningful spatial patterns. Conversely, methods that prioritize sparsity and feature selection (Tibshirani, 1996; Fan and Li, 2001; Candes and Tao, 2007) can lead to fragmented or noisy estimates, failing to preserve the spatial coherence that is inherent in many scientific imaging problems. These limitations underscore the need for methodological developments that can seamlessly integrate sparsity and smoothness and thereby produce interpretable and stable estimates of the underlying coefficient function.

In the literature of scalar-on-function regression, there exist methods that incorporate both localized sparsity and smoothness of the coefficient functions, such as James et al. (2009); Zhou et al. (2013); Lin et al. (2017); Gurer et al. (2024) and approaches based on the fused lasso framework (Tibshirani et al., 2005; Friedman et al., 2007). Relatedly, Xu et al. (2022) studied a locally sparse functional logistic regression model, extending the idea of localized sparsity to a functional generalized linear model setting. Despite these advances, the above methods are primarily developed for univariate functional predictors and do not directly accommodate bivariate or multivariate functional covariates, where one must simultaneously enforce smoothness along multiple spatial directions while maintaining localized sparsity. This gap motivates the development of smooth and sparse regression methods for scalar-on-image problems.

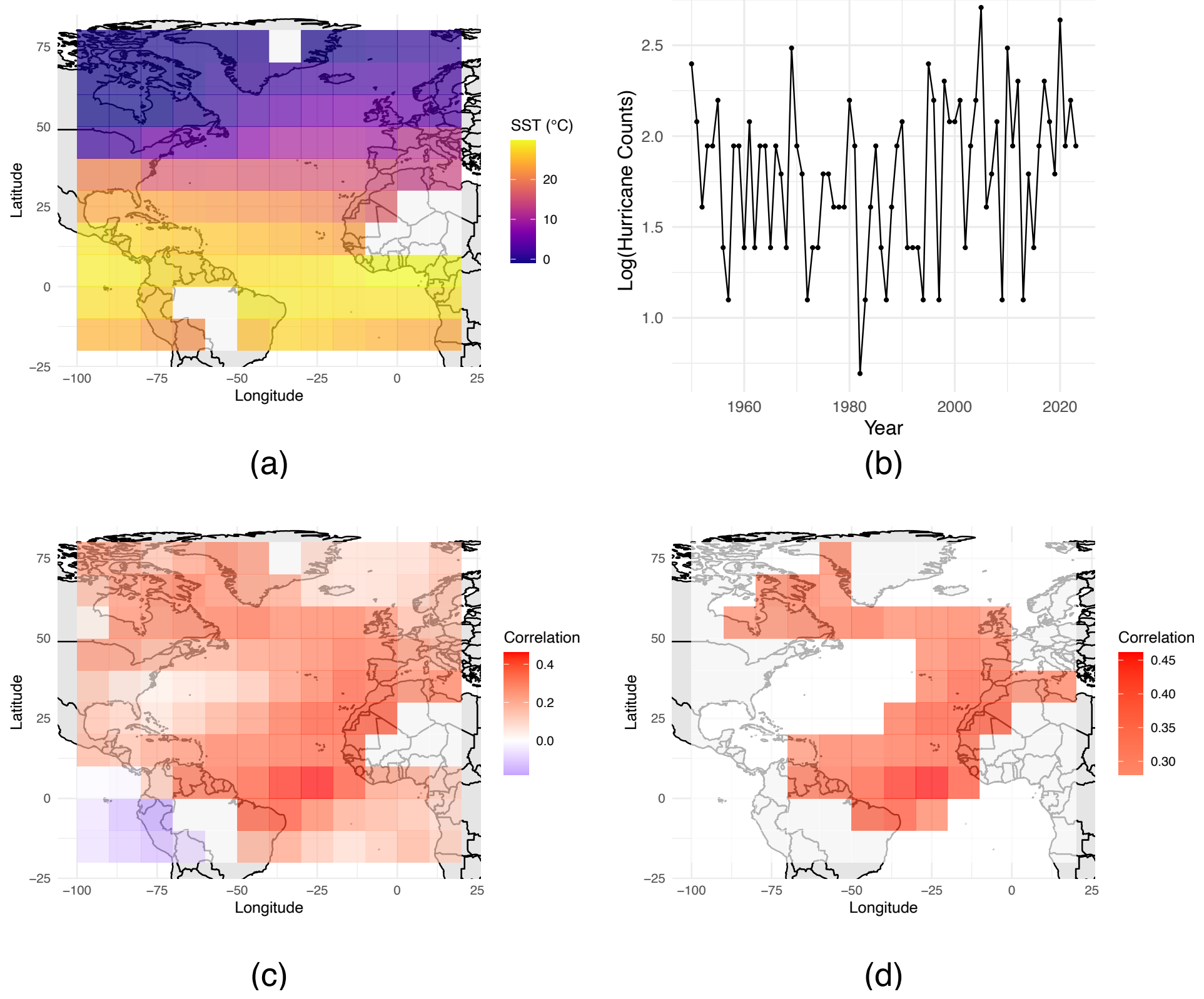

Figure 1: (a) March–April–May averaged Sea Surface Temperature (MAM-SST) in 2023; (b) logarithm of hurricane counts in the North Atlantic basin from 1950 to 2023; (c) correlations between MAM-SST and log(Hurricane Counts) over the same period; and (d) statistically significant correlations identified via a Student's $t$-test with Benjamini–Hochberg adjusted $p$-values at the 0.05 level. The region of interest is $\Omega = [20°\text{S}, 80°\text{N}] \times [100°\text{W}, 20°\text{E}]$.

In addition, approaches grounded in Bayesian variable selection (Goldsmith et al., 2014; Li et al., 2015; Kang et al., 2018), while effective in enhancing model interpretability, often entail substantial computational costs and require extensive hyperparameter tuning to achieve satisfactory performance. These limitations can hinder their practical applicability, especially in high-dimensional settings involving complex image data, where scalability and efficiency are critical considerations.

To facilitate interpretation and address existing limitations, we present a new class of scalar-on-image linear models that relate a scalar response to a two-dimensional functional covariate. The new generalizable method extends the univariate functional linear regression to the bivariate or even higher-dimensional settings. Such an extension is nontrivial, as it necessitates a unified framework to simultaneously enforcing smoothness along multiple spatial dimensions. Beyond the extension to bivariate setting, our key theoretical advancement is a general Generalized Dantzig Selector (GDS) analysis that allows the penalty operator to be a non-invertible linear transformation of basis coefficients, enabling the simultaneous incorporation of multi-directional derivative-based smoothness constraints and sparsity within a single convex program. The main theoretical result of this paper establishes finite-sample (non-asymptotic) upper bounds for (i) the $L_2$ estimation error of the coefficient surface, and (ii) the corresponding prediction error measured in the empirical $n$-norm, under mild regularity conditions formulated in terms of the GDS design.

A notable study by Wang et al. (2017) employs total variation regularization to enforce smoothness while preserving edges in image data. Their approach is limited to using piecewise constant functions to approximate the coefficient surface, with smoothness imposed through bounded total variation. In contrast, our proposed method offers greater flexibility by using various basis representations such as B-splines and incorporating higher-order difference operators to impose different degrees of smoothness. This additional flexibility enables more nuanced control over the smoothness and sparsity of the estimated coefficient surface, facilitating better adaptation to the underlying structural characteristics of image predictors.

The remainder of this paper is organized as follows. Section 2 introduces the necessary notation and presents the methodology underlying our proposed approach. Section 3 provides a theoretical analysis, establishing non-asymptotic error bounds for the estimation of the proposed method. In Section 4, we report simulation results under various settings and compare the proposed method with alternative approaches, including P-splines and functional principal component regression (FPCR). Section 5 illustrates the application of our method to a real-world dataset, where the goal is to predict hurricane counts using the sea surface temperature data. Finally, Section 6 discusses the limitations of the current work and suggests potential directions for future research.

# 2 METHODOLOGY

## 2.1 Notations

Let $\mathcal{D}_t$ and $\mathcal{D}_s$ denote the partial derivative operators with respect to $t$ and $s$, respectively. For $\alpha_i \in \mathbb{N}$ $(i = 1, 2)$, define $\mathcal{D}^{\alpha_1,\alpha_2} = \mathcal{D}_t^{\alpha_1}\mathcal{D}_s^{\alpha_2}$.

Let $L_2(\Omega)$ denote as the space of square-integrable functions defined on a domain $\Omega$, i.e.,

$$L_2(\Omega) = \left\{ f : \iint_\Omega f^2(t,s)\,dt\,ds < \infty \right\}, \text{and the norm } \|f\|_{L_2(\Omega)} = \left( \iint_\Omega f^2(t,s)\,dt\,ds \right)^{1/2}.$$

Functions $f, g \in L_2(\Omega)$ are equivalent in $L_2(\Omega)$ if $\|f - g\|_{L_2(\Omega)} = 0$.

For $d_i \in \mathbb{N}$ $(i = 1, 2)$, define

$$L_2^{d_1,d_2}(\Omega) = \left\{ f : \mathcal{D}^{\alpha_1,\alpha_2} f \in L_2(\Omega) \text{ for all } 0 \le \alpha_1 \le d_1,\ 0 \le \alpha_2 \le d_2 \right\}.$$

Vectors are denoted by lowercase letters and matrices by uppercase letters, with dimensions stated at first use. When no dimension is given, a lowercase symbol is interpreted as a scalar parameter and an uppercase symbol as a scalar constant (if not specified otherwise).

For $y = (y_1, ..., y_n)^\top \in \mathbb{R}^n$, define the $n$-norm

$$\|y\|_n^2 = \frac{1}{n}\sum_{i=1}^{n} y_i^2.$$

Given $X \in \mathbb{R}^{n\times p}$, and $X_{\cdot j}, j = 1, ..., p$ the $j$th column of matrix $X$. Let

$$N = \mathrm{diag}\left(\|X_{\cdot 1}\|_n, \ldots, \|X_{\cdot p}\|_n\right) \in \mathbb{R}^{p\times p}$$

be the diagonal matrix of column $n$-norms of $X$. Finally, for $v = (v_1, ..., v_p)^\top \in \mathbb{R}^p$, define

$$\|v\|_{\ell_1} = \sum_{j=1}^{p} |v_j|, \quad \|v\|_{\ell_2} = \left(\sum_{j=1}^{p} |v_j|^2\right)^{1/2}, \quad \|v\|_{\ell_\infty} = \max_{1\le j\le p} |v_j|.$$

## 2.2 Basis Representation

Throughout, we assume that bivariate coefficient function $\beta(t, s)$ satisfies:

- **Sparsity**: There exists a subset $\Omega_0 \subset \Omega$ with $\iint_{\Omega_0} dt\, ds > 0$ such that $\beta(t, s) = 0$, for all $(t, s) \in \Omega_0$.
- **Smoothness**: There exist $d_1, d_2 \in \mathbb{N}$ such that $\beta \in L_2^{d_1,d_2}(\Omega)$.

Given a set of basis functions $\{b_1(t, s), \ldots, b_p(t, s)\} \subset L_2(\Omega)$, let $\mathcal{V}_p$ denote their span. Then $\beta(t, s)$ admits the decomposition

$$\beta(t, s) = b^\top(t, s)\eta + e(t, s), \tag{2}$$

where $b(t, s) = [b_1(t, s), \ldots, b_p(t, s)]^\top \in \mathbb{R}^p$, $\eta \in \mathbb{R}^p$ is the coefficient vector, and $e(t, s)$ captures the approximation error outside $\mathcal{V}_p$. Define

$$\eta^* = \arg\min_{\eta\in\mathbb{R}^p} \left(\iint_\Omega \left(\beta(t, s) - b^\top(t, s)\eta\right)^2 dt\, ds\right)^{1/2}.$$

Then $\beta(t, s) - b^\top(t, s)\eta^*$ lies in the orthogonal complement of $\mathcal{V}_p$, and $\eta^*$ minimizes the $L_2$ approximation error. For further discussion, let $\beta^*(t, s) = b^\top(t, s)\eta^*$ and $\mathcal{E}_b = \|\beta - \beta^*\|_{L_2(\Omega)}$. We assume that $\beta^*(t, s)$ inherits the sparsity and smoothness of $\beta(t, s)$ under the chosen basis.

A widely used basis for constructing smooth surfaces in multivariate settings is the tensor product B-splines (Schumaker, 2007). This involves selecting knots and spline orders in each dimension to flexibly model underlying structures while mitigating overfitting. Let $\{\phi_k(t)\}_{k=1}^{p_1}$ and $\{\psi_l(s)\}_{l=1}^{p_2}$ be B-spline bases of orders $d_1$ and $d_2$ with simple knots $0 < t_1 < \cdots < t_{p_1-d_1} < 1$ and $0 < s_1 < \cdots < s_{p_2-d_2} < 1$, respectively, where $d_i \in \mathbb{N}$, $p_i \in \mathbb{N}^+$ and $p_i > d_i$ for $i = 1, 2$. A typical tensor product B-spline basis function is

$$b_j(t, s) = \phi_k(t)\psi_l(s), \quad j = (k-1)p_2 + l. \tag{3}$$

Define the maximal knot spacings

$$\Delta_1 = \max_{1\leq i\leq p_1-d_1-1} |t_i - t_{i+1}|, \quad \Delta_2 = \max_{1\leq i\leq p_2-d_2-1} |s_i - s_{i+1}|.$$

Theorem 12.7 of Schumaker (2007) (Chapter 12, page 491) shows that there exists $C > 0$ depending only on $d_1$ and $d_2$ such that

$$\mathcal{E}_b \leq C \left( \Delta_1^{d_1} \|\mathcal{D}_t^{d_1}\beta\|_{L_2(\Omega)} + \Delta_2^{d_2} \|\mathcal{D}_s^{d_2}\beta\|_{L_2(\Omega)} \right)$$

for all $\beta(t, s) \in L_2^{d_1,d_2}(\Omega)$. A general choice is the tensor product cubic B-splines, with $d_1 = d_2 = 4$. Another choice is the piecewise constant basis, which is a special case of the tensor product B-splines with $d_1 = d_2 = 1$, valued for its effectiveness in segmenting the input space into regions with approximately homogeneous properties. By increasing the number of pieces and incorporating smoothness regulation, one can approximate smooth surfaces while preserving simplicity. Suppose $p_1, p_2 \in \mathbb{N}^+$. Let $J_{t_k} = [t_k, t_{k+1})$ for $k = 1, \ldots, p_1 - 1$ and $J_{t_{p_1}} = [t_{p_1}, t_{p_1+1}]$, where $0 = t_1 < \cdots < t_{p_1+1} = 1$, and similarly define $J_{s_l} = [s_l, s_{l+1})$ for $l = 1, \ldots, p_2 - 1$ and $J_{s_{p_2}} = [s_{p_2}, s_{p_2+1}]$, with $0 = s_1 < \cdots < s_{p_2+1} = 1$. A piecewise constant basis function is

$$b_j(t, s) = \mathbf{1}\{t \in J_{t_k},\, s \in J_{s_l}\}, \tag{4}$$

where $j = (k-1)p_2 + l$, $k = 1, \ldots, p_1$, and $l = 1, \ldots, p_2$. It can be shown that for evenly spaced $t_1, ..., t_{p_1+1}$ and $s_1, ..., s_{p_2+1}$ in $[0, 1]$,

$$\mathcal{E}_b \lesssim \frac{1}{\sqrt{p}},$$

for all $\beta(t, s) \in L_2(\Omega)$, where $p = p_1 p_2$ is the total number of pieces.

Define

$$X_{ij} = \iint_\Omega x_i(t, s) b_j(t, s)\, dt\, ds, \quad \epsilon_i^* = \iint_\Omega x_i(t, s) e(t, s)\, dt\, ds + \epsilon_i, \quad i = 1, \ldots, n,$$

so that (1) can be written as $y = \alpha 1_n + X\eta + \epsilon^*$, where $y = (y_1, ..., y_n)$, $1_n \in \mathbb{R}^n$ is the vector of ones and $\epsilon^* = (\epsilon_1^*, ..., \epsilon_n^*)^\top$. Let

$$\bar{y} = \frac{1}{n}\sum_{i=1}^n y_i, \quad \bar{X} = \frac{1}{n}\sum_{i=1}^n X_{i\cdot}, \quad \bar{\epsilon}^* = \frac{1}{n}\sum_{i=1}^n \epsilon_i^*,$$

where $X_{i\cdot}$ denotes the $i$-th row of $X$, and define

$$\tilde{y} = y - \bar{y}1_n, \quad \tilde{X} = X - 1_n\bar{X}, \quad \tilde{\epsilon}^* = \epsilon^* - \bar{\epsilon}^* 1_n.$$

The model can then be centered as $\tilde{y} = \tilde{X}\eta + \tilde{\epsilon}^*$.

While one might consider applying variable selection methods to estimate $\eta$, sparsity on $\eta$ generally does not hold for arbitrary bases $b^\top(t, s)$, even if $\beta(t, s)$ is sparse. In Section 2.3, a generalized Dantzig selector is introduced to recover the sparsity and smoothness of $\beta(t, s)$. Once an estimator $\hat{\eta}$ is obtained, the intercept and coefficient function are estimated by

$$\hat{\alpha} = \bar{y} - \bar{X}\hat{\eta}, \quad \hat{\beta}(t, s) = b^\top(t, s)\hat{\eta}.$$

Without loss of generality, we henceforth assume the centered model

$$y = X\eta + \epsilon^*. \tag{5}$$

## 2.3 Proposal: The Generalized Dantzig Selector

The Dantzig selector (Candes and Tao, 2007) estimates sparse coefficients by minimizing the $\ell_1$ norm under an $\ell_\infty$ norm constraint on residual correlations. It behaves similarly to the Lasso (Tibshirani (1996); see also Bickel et al. (2009)) and can be efficiently solved via linear programming.

The FLiRTI method (James et al., 2009) applies the Dantzig selector to finite difference approximations of derivatives in univariate functional regression, relying on an invertible transformation matrix. Our method generalizes FLiRTI to the bivariate setting by introducing a generalized Dantzig selector based on bivariate difference operators, avoiding requiring an invertible transformation matrix.

Let $I_m$ denote the $m \times m$ identity matrix. The $d$th order difference matrix for $v \in \mathbb{R}^m$ is a $(m-d) \times m$ matrix, denoted by $D_m^d$, with entries

$$D_m^d(i,j) = \begin{cases} (-1)^k \binom{d}{k}, & \text{if } j = i + k, \quad k = 0, \ldots, d, \\ 0, & \text{otherwise.} \end{cases}$$

For example, when $d = 2$ and $m = 5$,

$$D_5^2 = \begin{pmatrix} 1 & -2 & 1 & 0 & 0 \\ 0 & 1 & -2 & 1 & 0 \\ 0 & 0 & 1 & -2 & 1 \end{pmatrix}.$$

By convention, $D_m^0 = I_m$.

Let $\mathcal{I} = \{(t_k, s_l) \in \Omega : 1 \leq k \leq m_1, 1 \leq l \leq m_2\}$ denote an evenly spaced grid, where a natural choice is the set of all image pixels. Define the bivariate difference operator

$$A^{d_1,d_2} = \left(\frac{1}{\delta_1}\right)^{d_1} \left(\frac{1}{\delta_2}\right)^{d_2} D_{m_1}^{d_1} \otimes D_{m_2}^{d_2},$$

where $\delta_1, \delta_2$ are the grid lengths in the $t$ and $s$ directions and $\otimes$ denotes the Kronecker product.

Our generalized Dantzig selector solves

$$
\begin{aligned}
&\min_{\eta \in \mathbb{R}^p} \|A\eta\|_{\ell_1}, \\
\text{subject to} \quad &\left\| \frac{1}{n} N^{-1} X^\top (y - X\eta) \right\|_{\ell_\infty} \leq \lambda,
\end{aligned} \tag{6}
$$

for some $\lambda > 0$, where

$$
A = \begin{bmatrix} wI_{m_1 m_2} \\ A^{d_1,d_2} \end{bmatrix} \Phi^\top \in \mathbb{R}^{L \times p}, \quad L = m_1 m_2 + (m_1 - d_1)(m_2 - d_2),
$$

$w > 0$ controls the sparsity penalty, and $\Phi$ is the matrix of basis function evaluations at points in $\mathcal{I}$ such that:

$$
\Phi^\top = \begin{bmatrix} b_1(t_1, s_1) & \cdots & b_p(t_1, s_1) \\ \vdots & & \vdots \\ b_1(t_{m_1}, s_{m_2}) & \cdots & b_p(t_{m_1}, s_{m_2}) \end{bmatrix}. \tag{7}
$$

To separately control smoothness along $t$ and $s$, we extend $A$ to

$$
A = \begin{bmatrix} wI_{m_1 m_2} \\ A^{d_1,0} \\ A^{0,d_2} \end{bmatrix} \Phi^\top \in \mathbb{R}^{L \times p}, \tag{8}
$$

where $L = m_1 m_2 + (m_1 - d_1) m_2 + m_1 (m_2 - d_2)$, $A^{d_1,0}$ and $A^{0,d_2}$ are partial difference operators along $t$ and $s$, respectively. The transformation $A$ links $\ell_1$ minimization with joint sparsity and smoothness enforcement in $\beta(t, s)$. Defining $\gamma = A\eta$, $\gamma$ approximates a concatenation of $w\beta(t, s)$, $\mathcal{D}_t^{d_1} \beta(t, s)$, and $\mathcal{D}_s^{d_2} \beta(t, s)$ evaluated at $\mathcal{I}$.

We adopt the Dantzig-type estimator here because it interfaces naturally with the transformation-based formulation used to encode both sparsity and multidirectional smoothness. Specifically, our estimator minimizes an $\ell_1$ norm of the transformed coefficient vector while imposing a constraint that controls the maximal correlation between the predictors and the residuals. This leads to a convex optimization problem that remains well-defined even when the transformation is not invertible. Alternative sparsity-inducing approaches are possible, such as Lasso-type penalization (e.g., Tibshirani et al. (2005) and Tibshirani and Taylor (2011)) or SCAD (Fan and Li, 2001). For example, a possible Lasso-type formulation under our setting will be

$$\min_{\eta \in \mathbb{R}^p} \|y - X\eta\|_{\ell_2}^2 + \lambda \|A\eta\|_{\ell_1}.$$

However, incorporating these alternatives would require additional tuning-parameter calibration that couples $\lambda$ with $A$ and/or nonconvex optimization analysis in SCAD, which is beyond the scope of this paper.

## 2.4 Computing Algorithm

Let $\gamma_j^+ = \max\{0, \gamma_j\}$ and $\gamma_j^- = \max\{0, -\gamma_j\}$ for $j = 1, \ldots, L$, and similarly define $\eta_j^+$ and $\eta_j^-$ for $j = 1, \ldots, p$. The computation proceeds as follows:

1. Given a grid $\mathcal{I}$ and data $\{(x_i(t, s), y_i) : (t, s) \in \mathcal{I}, i = 1, \ldots, n\}$, along with a basis matrix $\Phi$, orders $d_1, d_2$, and sparsity weight $w$, construct the design matrix $X$ and the transformation matrix $A$.

2. Center $X$ and $y$, and compute $N$, the diagonal matrix of column $n$-norms of the centered $X$.

3. For a given $\lambda > 0$, solve the following linear program:

$$\min_{\gamma^+,\gamma^-,\eta^+,\eta^-} \sum_{j=1}^{L}(\gamma_j^+ + \gamma_j^-)$$

$$\begin{aligned}\text{subject to} \quad & A(\eta^+ - \eta^-) = \gamma^+ - \gamma^-, \\ & X^\top X(\eta^+ - \eta^-) \leq n\lambda N + X^\top y, \\ & -X^\top X(\eta^+ - \eta^-) \leq n\lambda N - X^\top y, \\ & \gamma^+, \gamma^-, \eta^+, \eta^- \geq 0,\end{aligned}$$

where all inequalities are interpreted elementwise.

The optimization can be efficiently solved using standard linear programming packages, such as *lpSolve*(Berkelaar, 2024) in *R*.

# 3 THEORETICAL ANALYSIS

## 3.1 Preliminaries

We begin by introducing the necessary notations and definitions.

Define $\mu(x_i) = \iint_\Omega x_i(t,s)\beta(t,s)\,dt\,ds$ for $i = 1,\ldots,n$, and let $\mu = (\mu(x_1),\ldots,\mu(x_n))^\top$. Similarly, define $\hat{\mu}(x_i) = \iint_\Omega x_i(t,s)\hat{\beta}(t,s)\,dt\,ds$ and $\hat{\mu} = (\hat{\mu}(x_1),\ldots,\hat{\mu}(x_n))^\top$.

In the proposed method, the matrix $A$ defined in (8) is constructed to have full column rank ($L \geq p$). Suppose $A$ admits the singular value decomposition (SVD)

$$A = U_1 \begin{bmatrix} \Sigma \\ 0 \end{bmatrix} U_2^\top,$$

where $\Sigma$ is a $p \times p$ diagonal matrix with singular values in descending order, $U_1$ and $U_2$ are unitary matrices. Let

$$A^+ = U_2[\Sigma^{-1} \,|\, 0]U_1^\top$$

denote the Moore-Penrose pseudoinverse of $A$, satisfying $A^+A = I_p$. Define $V = XA^+$, so that (5) becomes

$$y = V\gamma + \epsilon^*.$$

Let $\mathcal{T} \subseteq \{1, \dots, L\}$ be an index set, and let $V_{\mathcal{T}}$ denote the submatrix of $V$ consisting of the columns indexed by $\mathcal{T}$. We introduce the following definitions related to the identifiability of $\gamma$ through $\ell_1$-minimization and the restricted eigenvalue assumptions, following Candes and Tao (2007) and Bickel et al. (2009).

**Definition 1** ($S$-restricted Minimum Eigenvalue)**.** For some integer S such that $1 \leq S \leq L$, the $S$-restricted minimum eigenvalue $\phi_S$ is the largest constant such that

$$\phi_S \|c\|_{\ell_2}^2 \leq \frac{\|V_{\mathcal{T}} c\|_{\ell_2}^2}{n}$$

for all $\mathcal{T}$ with $|\mathcal{T}| \leq S$ and all $c \in \mathbb{R}^{|\mathcal{T}|}$.

**Definition 2** ($S, S'$-restricted Orthogonality Constant)**.** For integers $S, S'$ such that $S+S' \leq L$, the $S, S'$-restricted orthogonality constant $\theta_{S,S'} \geq 0$ is the smallest constant such that

$$\frac{|\langle V_{\mathcal{T}} c, V_{\mathcal{T}'} c' \rangle|}{n} \leq \theta_{S,S'} \|c\|_{\ell_2} \|c'\|_{\ell_2}$$

for all disjoint $\mathcal{T}, \mathcal{T}'$ with $|\mathcal{T}| \leq S, |\mathcal{T}'| \leq S'$ and $c \in \mathbb{R}^{|\mathcal{T}|}$, $c' \in \mathbb{R}^{|\mathcal{T}'|}$.

A positive $\phi_S$ implies that $V_{\mathcal{T}}$ defines a bijection from $\mathbb{R}^{|\mathcal{T}|}$ onto its column space for any $|\mathcal{T}| \leq S$. A small $\theta_{S,S'}$ indicates that disjoint subsets of predictors are nearly orthogonal.

For $h \in \mathbb{R}^L$, let $h_{\mathcal{T}}$ denote both the subvector indexed by $\mathcal{T}$ and, by slight abuse of notation, the vector in $\mathbb{R}^L$ retaining entries in $\mathcal{T}$ and setting others to zero.

**Definition 3** (Restricted Eigenvalue I)**.** For some integer $S$ such that $1 \leq S \leq L$, define

$$\kappa_1(S) := \min_{\substack{\mathcal{T}_0 \subset \{1,\dots,L\} \\ |\mathcal{T}_0| \leq S}} \min_{\substack{h \neq 0 \\ \|h_{\mathcal{T}_0^{\complement}}\|_{\ell_1} \leq \|h_{\mathcal{T}_0}\|_{\ell_1}}} \frac{\|Vh\|_{\ell_2}}{\sqrt{n}\|h_{\mathcal{T}_0}\|_{\ell_2}}.$$

**Definition 4** (Restricted Eigenvalue II)**.** For integers $S, S'$ satisfying $1 \le S \le S' \le L$ and $S + S' \le L$, define

$$\kappa_2(S, S') := \min_{\substack{\mathcal{T}_0 \subset \{1,\dots,L\} \\ |\mathcal{T}_0| \le S}} \quad \min_{\substack{h \neq 0 \\ \|h_{\mathcal{T}_0^\complement}\|_{\ell_1} \le \|h_{\mathcal{T}_0}\|_{\ell_1}}} \frac{\|Vh\|_{\ell_2}}{\sqrt{n}\|h_{\mathcal{T}_{01}}\|_{\ell_2}},$$

where $\mathcal{T}_1$ is the index set corresponding to the $S'$ largest (in magnitude) entries of $h$ outside $\mathcal{T}_0$, and $\mathcal{T}_{01} = \mathcal{T}_0 \cup \mathcal{T}_1$.

## 3.2 Non-asymptotic Error Bound

In this section, we analyze the non-asymptotic $L_2$ error bound between $\hat{\beta}(t, s)$ and the true $\beta(t, s)$, as well as the $n$-norm error bound between $\hat{\mu}$ and $\mu$.

For a given basis $b(t, s) = [b_1(t, s), \dots, b_p(t, s)]^\top$, let $\eta^*$ and $\mathcal{E}_b$ be defined as before, and define $\gamma^* = A\eta^*$. Below, we list the technical assumptions used in the subsequent discussions.

**(A1)** The errors are independent Gaussian random variables:

$$\epsilon_i \sim \mathcal{N}(0, \sigma^2), \quad \text{independently for } i = 1, \dots, n, \text{ with } \sigma > 0.$$

**(A2)** The functional covariates are uniformly bounded in the $L_2(\Omega)$ norm:

$$\|x_i\|_{L_2(\Omega)} \le M < \infty, \quad \text{for all } i = 1, \dots, n.$$

**(A3)** The coefficient vector $\gamma^* \in \mathbb{R}^L$ has at most $S$ nonzero components.

**(A4)** The matrix $V$ satisfies $\phi_{2S} - \theta_{S,2S} > 0$.

We are now ready to present the main results of the paper.

**Theorem 1.** *Assume* ***(A1)*–*(A4)*** *hold. Let $\hat{\eta}$ be the solution to (6), and define $\hat{\gamma} = A\hat{\eta}$. Suppose*

$$\lambda = C\sigma\sqrt{\frac{\log p}{n}} + M\mathcal{E}_b,$$

*for some constant $C > \sqrt{2}$. Then, with probability at least $1 - p^{1-C^2/2}$,*

$$\|\hat{\beta} - \beta\|_{L_2(\Omega)} \leq \frac{4C_b\sqrt{L}N_{\max}}{(\phi_{2S} - \theta_{S,2S})\sigma^2_{\min}(A)} \lambda\sqrt{S} + \mathcal{E}_b, \tag{9}$$

*and*

$$\|\hat{\mu} - \mu\|_n \leq \frac{4\sqrt{L}N_{\max}}{\sqrt{\phi_{2S} - \theta_{S,2S}}\,\sigma_{\min}(A)} \lambda\sqrt{S} + M\mathcal{E}_b, \tag{10}$$

*where $\hat{\beta}(t,s) = b^\top(t,s)\hat{\eta}$, $C_b = \left(\sum_{j=1}^p \iint_\Omega b_j^2(t,s)\,dt\,ds\right)^{1/2}$, $N_{\max} = \max_{j=1,\ldots,p} N_j$, and $\sigma_{\min}(A)$ is the smallest singular value of $A$.*

*Remark* (Comparison with Candes and Tao (2007)). The condition $1 - \delta_{2S} > \theta_{S,2S}$ used in Candes and Tao (2007) guarantees identifiability of $\gamma$, where $\delta_S$ is the $S$-restricted isometry constant (see also Candes and Tao (2005)). Specifically, $\delta_S$ is the smallest constant such that

$$(1 - \delta_S)\|c\|^2_{\ell_2} \leq \frac{\|V_\mathcal{T} c\|^2_{\ell_2}}{n} \leq (1 + \delta_S)\|c\|^2_{\ell_2}$$

for all $|\mathcal{T}| \leq S$. Here in **(A4)**, we adopt the alternative condition $\phi_{2S} > \theta_{S,2S}$, which avoids requiring an upper bound on the restricted eigenvalues of $V^\top V/n$ while still ensuring the proof to proceed.

However, the condition **(A4)** is generally difficult to verify and may not hold for arbitrary design matrices. In Theorem 2, we relax this requirement by adopting a restricted eigenvalue condition, following a similar approach to that in Bickel et al. (2009). Below, we list the restricted eigenvalue conditions.

**(A5)** The matrix $V$ satisfies the restricted eigenvalue condition $\kappa_1(S) > 0$.

**(A6)** The matrix $V$ satisfies the stronger restricted eigenvalue condition $\kappa_2(S, S) > 0$.

We note that condition **(A4)** serves as a sufficient condition for both **(A5)** and **(A6)**, and that **(A6)** further implies **(A5)**. For additional discussion, we refer the reader to the Supplementary Material and Bickel et al. (2009). Our main result, Theorem 2, requires only one of the weaker conditions—either **(A5)** or **(A6)**—leading to slightly different bounds depending on the assumption adopted.

**Theorem 2.** *Assume **(A1)**–**(A3)** and **(A5)** hold. Let $\hat{\eta}$ be the solution to (6), and define $\hat{\gamma} = A\hat{\eta}$. Suppose*

$$\lambda = C\sigma\sqrt{\frac{\log p}{n}} + M\mathcal{E}_b,$$

*for some constant $C > \sqrt{2}$. Then, with probability at least $1 - p^{1-C^2/2}$,*

$$\|\hat{\beta} - \beta\|_{L_2(\Omega)} \leq \frac{8C_b\sqrt{L}N_{\max}}{\kappa_1^2(S)\sigma_{\min}^2(A)}\,\lambda S + \mathcal{E}_b, \tag{11}$$

*and*

$$\|\hat{\mu} - \mu\|_n \leq \frac{4\sqrt{L}N_{\max}}{\kappa_1(S)\,\sigma_{\min}(A)}\,\lambda\sqrt{S} + M\mathcal{E}_b, \tag{12}$$

*with the same definitions of $\hat{\beta}(t,s)$, $C_b$, $N_{\max}$, and $\sigma_{\min}(A)$ as in Theorem 1.*

*If, in addition, **(A6)** holds, then with the same probability,*

$$\|\hat{\beta} - \beta\|_{L_2(\Omega)} \leq \frac{8C_b\sqrt{L}N_{\max}}{\kappa_2^2(S,S)\,\sigma_{\min}^2(A)}\,\lambda\sqrt{S} + \mathcal{E}_b. \tag{13}$$

*Remark* (On interpretation of the Bound)*.* Let $\beta^*(t,s) = b^\top(t,s)\eta^*$ denote the "oracle" approximation under basis $b$. Then

$$\|\hat{\beta} - \beta\|_{L_2(\Omega)} \leq \|\hat{\beta} - \beta^*\|_{L_2(\Omega)} + \|\beta^* - \beta\|_{L_2(\Omega)},$$

where $\|\beta^* - \beta\|_{L_2(\Omega)} = \mathcal{E}_b$ is fixed given $b$. The bound on $\|\hat{\beta} - \beta^*\|_{L_2(\Omega)}$ depends on $\lambda$, which itself depends on $\mathcal{E}_b$. In practice, to achieve good approximation, especially with piecewise constant bases, $p$ should be chosen large enough so that $M\mathcal{E}_b$ is dominated by $C\sigma\sqrt{(\log p)/n}$.

Several constants appearing in Theorems 1 and 2 merit further discussion. Among them, the quantities $\phi_{2S} - \theta_{S,2S}$, $\kappa_1(S)$, and $\kappa_2(S,S)$ depend on the sparsity level $S$, the design matrix $X$, and the transformation matrix $A$. However, these terms are generally intractable and cannot be computed in closed form. To provide some intuition, we include a simple illustrative example in the Supplementary Material where assumptions **(A5)** and **(A6)** are not numerically violated, and we provide the estimates of $\kappa_1(S)$ and $\kappa_2(S, S)$ in that setting.

In contrast, the other constants in the bounds are more transparent and can be readily computed. Specifically, $N_{\max}$ depends only on the covariates $X$, $C_b$ is determined solely by the choice of basis functions $b$, and both $L$ and $\sigma_{\min}(A)$ rely only on the transformation matrix $A$.

Given the basis matrix $\Phi^\top$ defined in (7), it is possible to choose the weight $w$ to make $\sqrt{L}/\sigma_{\min}(A)$ a constant. For example, using the piecewise constant basis defined in (4), and aligning the estimation grid in (8) with the partitioning (i.e., one grid point per piece), $\Phi^\top$ becomes the identity matrix, yielding $C_b = 1$ and $\sigma_{\min}(A) = w$. Moreover, setting $w = \sqrt{L}$ ensures $\sqrt{L}/\sigma_{\min}(A) = 1$, leading to the simplified bound in (13)

$$\|\hat{\beta} - \beta\|_{L_2(\Omega)} \leq \frac{8N_{\max}}{\kappa_2^2(S, S)} \lambda \sqrt{\frac{S}{L}} + \mathcal{E}_b.$$

Here, $S/L$ can be roughly interpreted as the average area proportion of the nonzero regions of $\beta(t, s)$, $\mathcal{D}_t^{d_1}\beta(t, s)$, and $\mathcal{D}_s^{d_2}\beta(t, s)$ within $\Omega$ (if the partial derivatives exist).

Although a larger $w$ increases $\sigma_{\min}(A)$, it does not necessarily improve estimation. On one hand, $w$ rescales $A$ and thus affects $\kappa_2(S, S)$. On the other hand, an extremely large $w$ may lead to an overly sparse estimate, sacrificing smoothness. Empirically, our estimator shows limited sensitivity to $w$ within a reasonable range. Accordingly, $w$ may be fixed at a positive constant for convenience, or tuned by cross-validation over a small grid scales with $\sqrt{L}$.

*Remark* (On sparsistency for $\text{supp}(A\eta^*)$). An appealing theoretical direction is to establish sparsistency for the proposed estimator (i.e., exact recovery of the nonzero pattern promoted by the penalty). Since our formulation induces sparsity through the objective $\|A\eta\|_{\ell_1}$, the natural sparsistency target is support recovery of the analysis coefficients $\gamma^* := A\eta^*$, namely

$$\mathbb{P}\{\mathrm{supp}(A\hat{\eta}) = \mathrm{supp}(A\eta^*)\} \to 1.$$

Such a result typically requires stronger ingredients than the $\ell_2$-type estimation error bounds developed in this paper. Notably, one may need (see, for example, Wainwright (2009); Zhao and Yu (2006)): (i) an $\ell_\infty$-type (or sign) control for $\hat{\gamma} - \gamma^*$, where $\hat{\gamma} := A\hat{\eta}$; (ii) a beta-min separation condition ensuring that $\min_{j \in \mathrm{supp}(\gamma^*)} |\gamma_j^*|$ dominates the $\ell_\infty$ error level; and (iii) an additional structural assumption preventing excessive dependence between active and inactive components (e.g., an irrepresentability or mutual incoherence condition formulated for the operator-induced representation together with the normalized design). In the current work, we provide an $\ell_2$ control for $\hat{\gamma} - \gamma^*$ (see Lemma 6 and Lemma 8 in the Supplementary Material), which implies the basic bound $\|\hat{\gamma} - \gamma^*\|_{\ell_\infty} \leq \|\hat{\gamma} - \gamma^*\|_{\ell_2}$; however, a sharper $\ell_\infty$ control is generally needed for exact support recovery. Verifying (iii) is also technically challenging in our setting because $A$ is generally non-invertible and combines an identity block with a difference block $A^{d_1,d_2}$, leading to substantial correlation among coordinates of $\gamma = A\eta$. We will leave the sparsistency theory for $\mathrm{supp}(\gamma^*)$ to future work.

# 4 SIMULATION STUDY

## 4.1 Evaluations

To evaluate the performance of the proposed method and competing approaches under different settings, we compute the Mean Squared Error (MSE)

$$\mathrm{MSE} = \frac{1}{n} \sum_{i=1}^{n} (\hat{\mu}(x_i) - y_i)^2$$

on an independent test set of size 10,000.

In addition to MSE, we consider the Relative Integrated Squared Error (RISE) of $\hat{\beta}$,

$$\text{RISE}(\hat{\beta}) = \frac{\iint_\Omega (\hat{\beta}(t,s) - \beta(t,s))^2 \, dt \, ds}{\iint_\Omega \beta^2(t,s) \, dt \, ds},$$

as well as the recovery rates of the zero and nonzero regions, defined respectively as

$$R_1(\hat{\beta}) = \frac{\iint_\Omega I_{\{|\hat{\beta}(t,s)|=0\} \cap \{|\beta(t,s)|=0\}}(t,s) \, dt \, ds}{\iint_\Omega I_{\{|\beta(t,s)|=0\}}(t,s) \, dt \, ds},$$

and

$$R_2(\hat{\beta}) = \frac{\iint_\Omega I_{\{|\hat{\beta}(t,s)|>0\} \cap \{|\beta(t,s)|>0\}}(t,s) \, dt \, ds}{\iint_\Omega I_{\{|\beta(t,s)|>0\}}(t,s) \, dt \, ds}.$$

These metrics collectively assess both the prediction accuracy and the ability to recover the true sparsity structure of the coefficient function $\beta(t,s)$.

## 4.2 Simulation Settings

In the simulations, data were generated from the following $\beta$ surfaces:

- $\beta_1(t,s) = \phi(t)\phi(s)$, where $\phi(t) = \sin(\pi t) - \sin(\pi/4)$ for $t \in [0, 1/4]$, $\phi(t) = \sin(3\pi/4) - \sin(\pi t)$ for $t \in [3/4, 1]$, and $\phi(t) = 0$ otherwise.

- $\beta_2(t,s) = \phi(t)\phi(s)$, where $\phi(t) = 100\left(\exp(t - 0.5)^2 - \exp(0.04)\right)$ for $t \in [0.3, 0.7]$, and $\phi(t) = 0$ otherwise.

- $\beta_3(t,s)$ is piecewise defined: $200\left((t-0.6)^2 + (s-0.4)^2 - 0.04\right)$ if $(t-0.6)^2 + (s-0.4)^2 \leq 0.04$; $200\left(0.04 - (t-0.4)^2 - (s-0.6)^2\right)$ if $(t-0.4)^2 + (s-0.6)^2 \leq 0.04$; and 0 otherwise.

The true $\beta$ surfaces are shown in the leftmost column of Figure 2.

Each predictor surface was generated as

$$x(t,s) = \sum_{k=1}^{m} \sum_{j=1}^{m} a_{jk} \psi_j(t) \psi_k(s),$$

where $a_{jk} \sim \mathcal{N}(0,1)$ independently for $j, k = 1, \ldots, m$, and $\{\psi_j(t)\}_{j=1}^m$ denotes a set of basis functions. We considered the following basis settings:

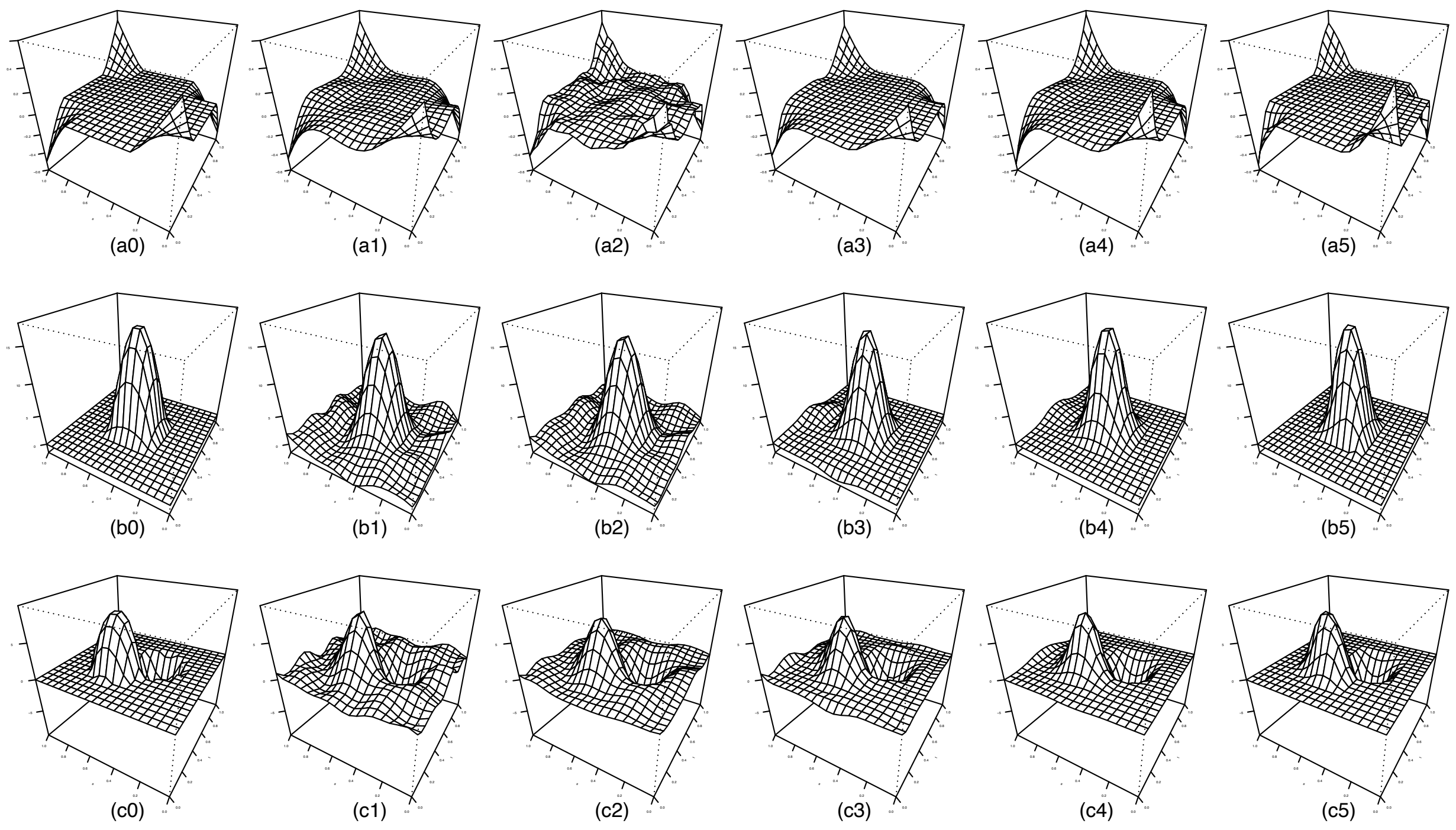

Figure 2: True coefficient surfaces $\beta$ (a0–c0) and fitted surfaces $\hat{\beta}$ under predictor $P_1$ by P-splines (a1–c1), FPCR (a2–c2), and the proposed methods: $\text{GDS}_{B_1}$ (a3–c3), $\text{GDS}_{B_2}$ (a4–c4), and $\text{GDS}_{B_2,\text{AIC},\text{Re}}$ (a5–c5).

$P_1$: B-splines of order 4 with 6 interior knots evenly spaced on $[0, 1]$, yielding $m = 10$.

$P_2$: B-splines of order 5 with 15 interior knots evenly spaced on $[0, 1]$, yielding $m = 20$.

All combinations $(\beta_j, P_k)$ for $j = 1, 2, 3$ and $k = 1, 2$ were evaluated.

Random noise was generated from $\mathcal{N}(0, \sigma^2)$ to attain a signal-to-noise ratio

$$\text{SNR}(\sigma) = \frac{\text{Var}\left(\iint_\Omega x_i(t,s)\beta(t,s)\,dt\,ds\right)}{\text{Var}(\epsilon_i)}$$

of 4. All integrals were approximated via Gaussian quadrature with equal weights at the grid points $\mathcal{I}$ with $m_1 \times m_2 = 400$. Experiments were conducted with a training sample size of $n = 400$ (unless specified otherwise), comparing the proposed method with the penalized B-splines (P-splines; see Wood (2017)) and functional principal components regression (FPCR; see Reiss and Ogden (2007)).

Let $B_1$ denote the basis defined in (3), with 7 interior knots and order 4 in each direction (tensor-product cubic B-spline), and $B_2$ denote the piecewise constant basis with $p_1 = p_2 = 20$ in (4). We denote the method using $B_1$ and $B_2$ as $\text{GDS}_{B_1}$ and $\text{GDS}_{B_2}$, respectively.

The smoothness orders $d_1$ and $d_2$ may be selected from a small candidate set (e.g., $\{0, 1, 2, 3\}$). While third-order differencing is less directly interpretable than first- or second-order differencing, allowing the option $d_1, d_2 = 3$ is useful in practice because it promotes piecewise–quadratic-like behavior on the grid by discouraging rapid changes in curvature. To reduce computational burden, the differencing orders were fixed in the main simulation: $d_1 = d_2 = 2$ for the tensor-product cubic B-spline basis $B_1$, and $d_1 = d_2 = 3$ for the piecewise-constant basis $B_2$. An illustrative experiment in the Supplementary Material demonstrates the selection of these orders. The remaining tuning parameters $\lambda$ and $w$ were chosen jointly by an independent validation set of size 400, using a grid search with candidates scales proportionally to $\sqrt{\log p/n}$ and $\sqrt{L}$, respectively.

The P-splines method was fitted using $B_1$, with the smoothing parameter chosen by minimizing the validation error. For FPCR, the *fcpr()* in the R package *refund* (Goldsmith et al., 2025) was used for implementation. The number of components was selected to explain $100\,pve\%$ of the predictor variation, where $pve \in \{0.90, 0.91, ..., 0.99\}$ was chosen by the same validation set. Additionally, $\text{GDS}_{B_2}$ was tuned using AIC, with the corresponding post-refitted results denoted as $\text{GDS}_{B_2,\text{AIC},\text{Re}}$ (see, e.g., Candes and Tao (2007)). Refitting was applied only when AIC was used, and only the post-refitted results are reported in Section 4.3. We refer to the Supplementary Material for additional details on parameters tuning and the refitting step.

## 4.3 Results

The models were fitted using the selected tuning parameters and the mean squared errors (MSEs) were evaluated on an independent test set. This procedure was repeated 100 times to compute the average MSEs and their standard errors, with results summarized in Table 1.

To further assess coefficient estimation and interpretability, we calculated the average relative integrated squared errors (RISEs) of $\hat{\beta}$, the average zero region recovery rates ($R_1(\hat{\beta})$), the average nonzero region recovery rates ($R_2(\hat{\beta})$), and their corresponding standard errors. These results are reported in Tables 2, 3, and 4.

| Method | $\beta_1$ ($10^{-6}$) | | $\beta_2$ ($10^{-2}$) | | $\beta_3$ ($10^{-2}$) | |
|---|---|---|---|---|---|---|
| | $P_1$ | $P_2$ | $P_1$ | $P_2$ | $P_1$ | $P_2$ |
| P-splines | 10.12 (0.03) | 5.18 (0.01) | 3.42 (0.01) | 0.99 (0.003) | 1.01 (0.003) | 0.37 (0.001) |
| FPCR | 10.68 (0.03) | 5.41 (0.02) | 3.35 (0.01) | 0.98 (0.003) | 0.98 (0.003) | 0.37 (0.001) |
| $\text{GDS}_{B_1}$ | **9.96 (0.03)** | **5.13 (0.01)** | 3.19 (0.01) | **0.90 (0.002)** | 0.94 (0.003) | **0.34 (0.001)** |
| $\text{GDS}_{B_2}$ | 9.99 (0.03) | 5.29 (0.02) | 3.18 (0.01) | 0.92 (0.003) | 0.93 (0.003) | 0.35 (0.001) |
| $\text{GDS}_{B_2,\text{AIC},\text{Re}}$ | **9.97 (0.03)** | 5.20 (0.02) | **3.13 (0.01)** | **0.89 (0.002)** | **0.91 (0.002)** | **0.33 (0.001)** |

Table 1: Mean squared errors (MSEs) (in indicated units) and standard errors (in parentheses) across methods and settings.

In terms of both prediction accuracy and $\beta$ estimation, the proposed methods $\text{GDS}_{B_1}$ and $\text{GDS}_{B_2}$ consistently outperform P-splines and FPCR across most settings. Although $\text{GDS}_{B_2,\text{AIC},\text{Re}}$ shows suboptimal performance under limited sample sizes for the $\beta_1$ setting, its performance improves substantially as the sample size increases, as illustrated in Figures 3 and 4.

The mean values of $R_1(\hat{\beta})$ and $R_2(\hat{\beta})$ in Tables 3 and 4 were computed by truncating $\hat{\beta}(t,s)$ to zero where its absolute value was smaller than $10^{-8}$. It is evident that the P-splines and FPCR methods fail to identify zero and nonzero regions under any setting. In contrast, the proposed method using the piecewise constant basis (GDS$_{B_2}$) identifies over 30% of the true zero regions while maintaining more than 99% precision in identifying nonzero regions for $\beta_2$ and $\beta_3$. Moreover, the AIC-tuned version (GDS$_{B_2,\text{AIC,Re}}$) successfully recovers over 89% of the true zero regions across all settings, while achieving over 91% precision for the nonzero regions.

| Method | $\beta_1$ | | $\beta_2$ | | $\beta_3$ | |
|---|---|---|---|---|---|---|
| | $P_1$ | $P_2$ | $P_1$ | $P_2$ | $P_1$ | $P_2$ |
| P-splines | 2.23 (0.04) | 2.25 (0.04) | 9.01 (0.16) | 5.58 (0.09) | 11.35 (0.17) | 8.18 (0.10) |
| FPCR | 6.79 (0.17) | 3.59 (0.07) | 7.29 (0.12) | 5.52 (0.09) | 8.16 (0.12) | 7.13 (0.09) |
| GDS$_{B_1}$ | **1.96 (0.05)** | **2.15 (0.05)** | 2.63 (0.09) | **1.70 (0.05)** | 4.60 (0.08) | 4.33 (0.06) |
| GDS$_{B_2}$ | 2.10 (0.05) | 3.02 (0.08) | 3.10 (0.10) | 2.92 (0.07) | 5.04 (0.12) | 5.09 (0.10) |
| GDS$_{B_2,\text{AIC,Re}}$ | 2.80 (0.15) | 3.03 (0.13) | **2.33 (0.17)** | **1.71 (0.09)** | **4.11 (0.10)** | **3.67 (0.09)** |

Table 2: Mean relative integrated squared errors (RISEs) (in units of $10^{-2}$) and standard errors (in parentheses) across methods and settings.

| Method | $\beta_1$ | | $\beta_2$ | | $\beta_3$ | |
|---|---|---|---|---|---|---|
| | $P_1$ | $P_2$ | $P_1$ | $P_2$ | $P_1$ | $P_2$ |
| P-splines | 0.00 (0.00) | 0.00 (0.00) | 0.00 (0.00) | 0.00 (0.00) | 0.00 (0.00) | 0.00 (0.00) |
| FPCR | 0.00 (0.00) | 0.00 (0.00) | 0.00 (0.00) | 0.00 (0.00) | 0.00 (0.00) | 0.00 (0.00) |
| GDS$_{B_1}$ | 5.25 (0.50) | 3.84 (0.41) | 23.16 (0.69) | 36.03 (1.57) | 26.20 (1.01) | 21.69 (0.89) |
| GDS$_{B_2}$ | 9.40 (1.31) | 5.31 (0.70) | 56.92 (2.24) | 44.24 (2.09) | 55.92 (2.02) | 36.42 (1.84) |
| GDS$_{B_2,\text{AIC,Re}}$ | **93.94 (0.45)** | **89.58 (0.77)** | **99.07 (0.15)** | **97.88 (0.25)** | **93.04 (0.25)** | **92.11 (0.37)** |

Table 3: Mean $R_1(\hat{\beta})$ values (percentage of correctly identified zero regions) and standard errors (in parentheses) across methods and settings.

Figure 2 displays the estimated $\beta$ surfaces from various methods under predictor $P_1$, alongside the true surface. While the P-splines and FPCR methods produce bumpy estimates, the proposed methods clearly recover the flat zero regions in $\beta_1$, $\beta_2$, and $\beta_3$, providing interpretable insights into the shape and localization of predictive versus non-predictive regions. These findings align with the theoretical results in Section 3 and demonstrate that the proposed method effectively yields smooth, interpretable coefficient surfaces while accurately distinguishing between relevant and irrelevant regions.

| Method | $\beta_1$ | | $\beta_2$ | | $\beta_3$ | |
|---|---|---|---|---|---|---|
| | $P_1$ | $P_2$ | $P_1$ | $P_2$ | $P_1$ | $P_2$ |
| P-splines | 100.00 (0.00) | 100.00 (0.00) | 100.00 (0.00) | 100.00 (0.00) | 100.00 (0.00) | 100.00 (0.00) |
| FPCR | 100.00 (0.00) | 100.00 (0.00) | 100.00 (0.00) | 100.00 (0.00) | 100.00 (0.00) | 100.00 (0.00) |
| $\text{GDS}_{B_1}$ | 99.89 (0.04) | 99.94 (0.02) | 99.97 (0.02) | 100.00 (0.00) | 99.99 (0.01) | 99.99 (0.01) |
| $\text{GDS}_{B_2}$ | 99.92 (0.03) | 99.82 (0.05) | 99.94 (0.03) | 99.77 (0.06) | 99.76 (0.09) | 99.88 (0.04) |
| $\text{GDS}_{B_2,\text{AIC,Re}}$ | 93.63 (0.52) | 91.44 (0.63) | 97.41 (0.74) | 98.33 (0.21) | 96.30 (0.41) | 95.91 (0.36) |

Table 4: Mean $R_2(\hat{\beta})$ values (percentage of correctly identified nonzero regions) and standard errors (in parentheses) across methods and settings.

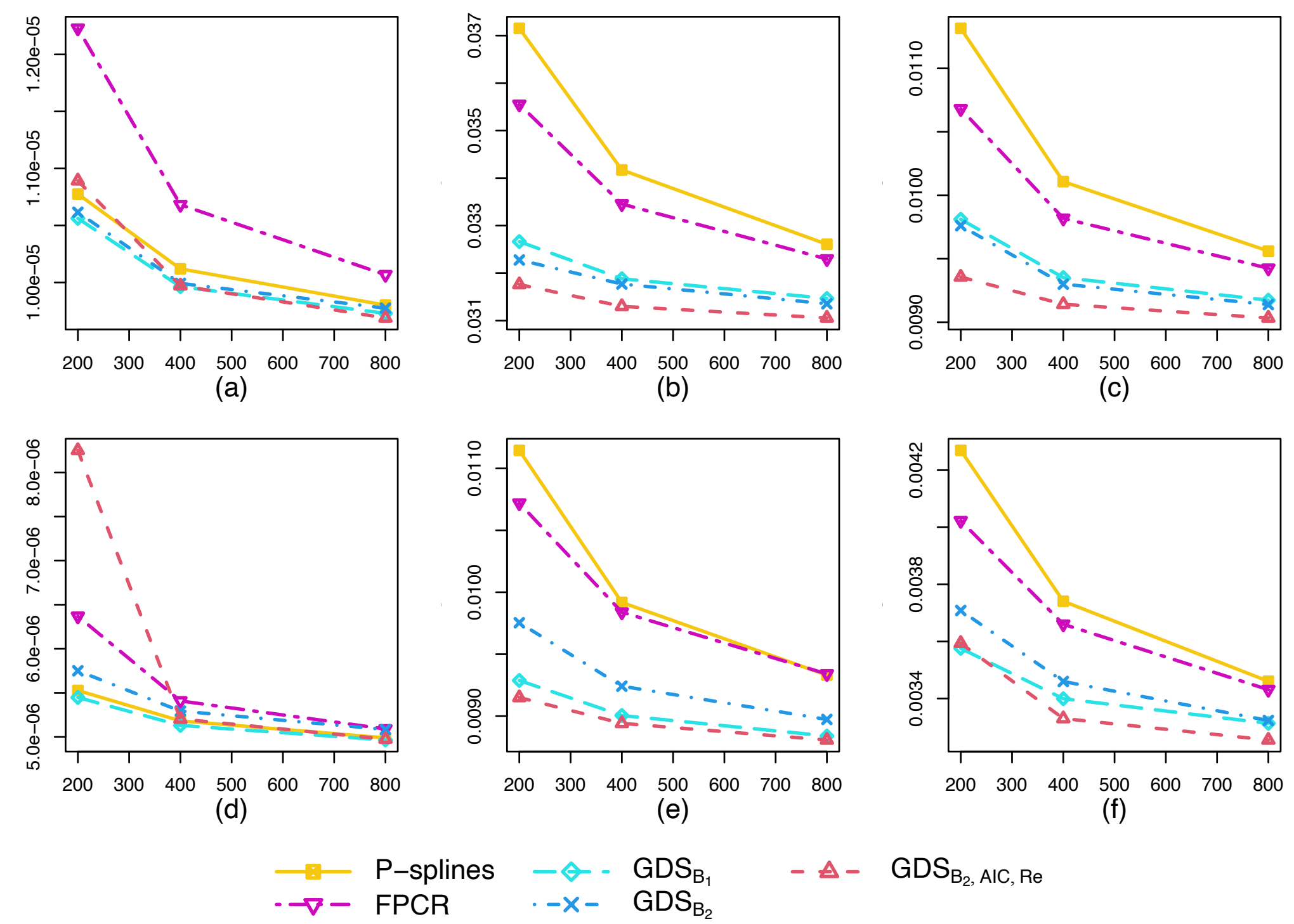

Figure 3: Mean squared errors (MSEs) versus sample size (200, 400, 800) across predictor settings ($P_1$: panels a–c; $P_2$: panels d–f) and coefficient settings ($\beta_1$: panels a, d; $\beta_2$: panels b, e; $\beta_3$: panels c, f).

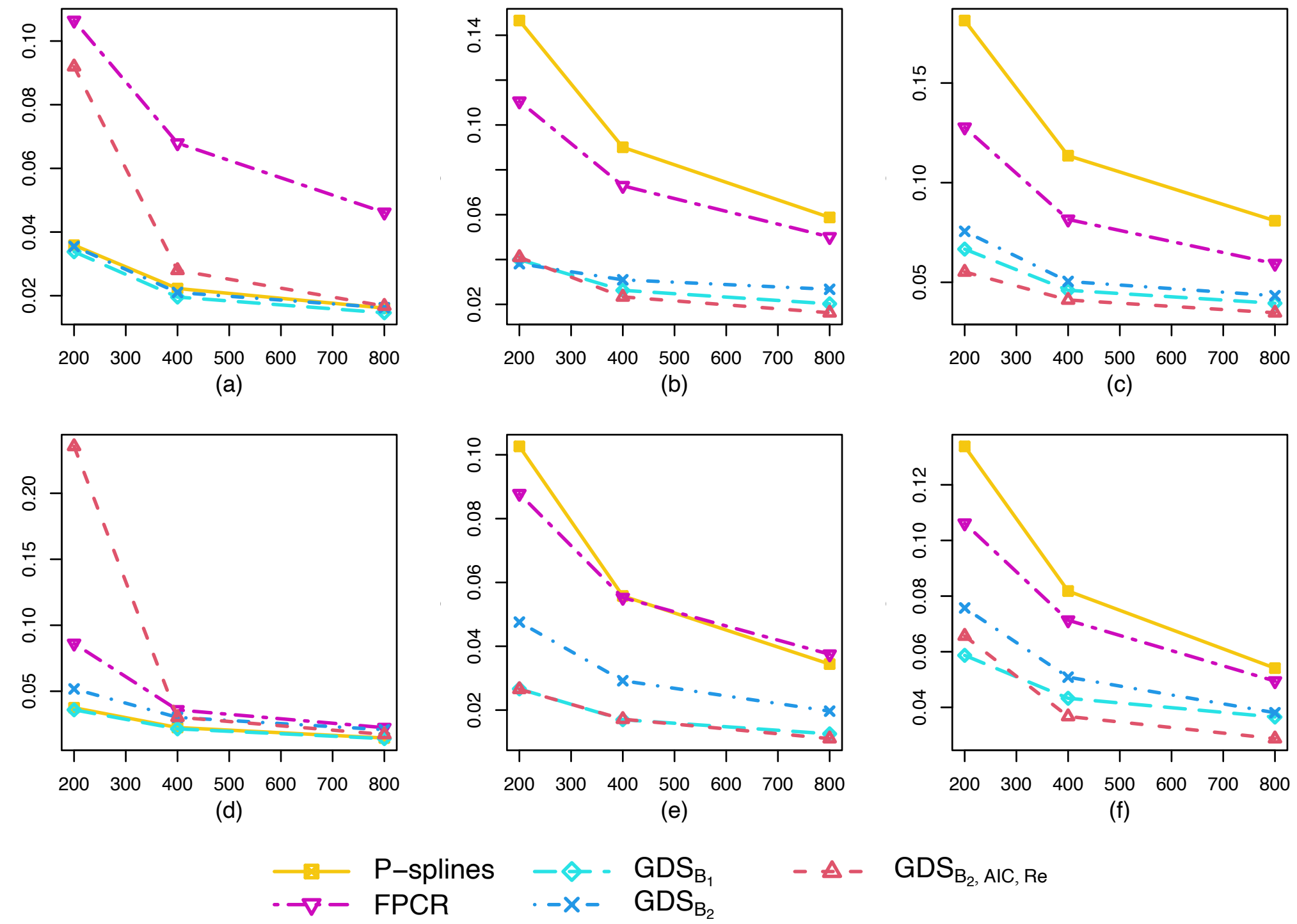

Figure 4: Mean relative integrated squared errors (RISEs) versus sample size (200, 400, 800) across predictor settings ($P_1$: panels a–c; $P_2$: panels d–f) and coefficient settings ($\beta_1$: panels a, d; $\beta_2$: panels b, e; $\beta_3$: panels c, f).

# 5 REAL DATA STUDY

Hurricane activity is widely recognized to be influenced by several large-scale environmental factors, including sea surface temperatures (SSTs), the El Niño–Southern Oscillation (ENSO), sea level pressure (SLP), the Atlantic Multidecadal Oscillation (AMO), and vertical wind shear, among others. Established frameworks for studying and predicting hurricane activity include models developed by NOAA's Climate Prediction Center (CPC) (Bell and Chelliah, 2006), the Tropical Storm Risk (TSR) consortium (Saunders and Lea, 2008), and Colorado

State University (CSU) (Klotzbach and Gray, 2009; Klotzbach et al., 2017). These models primarily focus on forecasting seasonal metrics, such as the total number of hurricanes. Among environmental predictors, SSTs are considered particularly critical, with many models (Bell and Chelliah, 2006; Vimont and Kossin, 2007; Saunders and Lea, 2008; Klotzbach and Gray, 2009; Davis et al., 2015) leveraging SSTs from subtropical or tropical regions to improve prediction accuracy.

The UA model proposed in Davis et al. (2015) employs models a linear relationship between the logarithm of hurricane counts and environmental predictors, including SSTs. Specifically, it uses the average SSTs from March to May within the region 0°–20°N and 60°W–10°E to construct a scalar SST predictor. This model provides a simple yet effective approach for forecasting hurricane counts in the North Atlantic Basin (NAB).

We extend the analysis by applying our proposed method to investigate the relationship between hurricane counts and SSTs in the NAB, while enhancing interpretability through spatial variability. To focus on SST effects, we consider a simplified model assuming consistent hurricane patterns over the years and excluding other predictors, with the region of interest defined as $\Omega = [20^\circ\mathrm{S}, 80^\circ\mathrm{N}] \times [100^\circ\mathrm{W}, 20^\circ\mathrm{E}]$.

The problem is formulated by the scalar-on-image model in (1), where $y_i$ denotes the logarithm of hurricane counts in year $i$, and $x_i(t, s)$ represents the March–April–May averaged SST (MAM-SST) at location $(t, s)$. Since SSTs are defined only over ocean surfaces, the integral is effectively restricted to the sea surface within $\Omega$. The values of $x_i(t, s)$ were set to zero over land, which does not affect the model, as land regions contribute nothing to the integral.

Sea surface temperature data were obtained from the Extended Reconstructed Sea Surface Temperature version 5 (ERSSTv5) dataset Huang et al. (2017), and historical hurricane counts for the NAB were accessed via the International Best Track Archive for Climate

Stewardship (IBTrACS) Knapp et al. (2010); Gahtan et al. (2024). Data from 1950 to 2004 were used for training, and data from 2005 to 2023 were reserved for testing, resulting in a training-to-testing split of approximately 3:1. For simplicity, the domain $\Omega$ was partitioned into $10^\circ \times 10^\circ$ grids, and the region-averaged MAM-SST values were used as predictors; see Figure 1. Integrals over $\Omega$ were computed using Gaussian quadrature, with weights proportional to the relative area of each region.

In the proposed method, the symmetric choice $d_1 = d_2 = d$ is imposed to reduce computational cost. We used $B_1$ (cubic B-splines with six degrees of freedom in each direction) and $B_2$ (with $p_1 = 12$ and $p_2 = 10$) for modeling. For the P-splines method, the same $B_1$ basis was employed. For the FPCR method, the number of basis functions in each direction was set to six, with the default penalty order of two. Model parameters, i.e., $(d, w, \lambda)$ for $\text{GDS}_{B_1}$ and $\text{GDS}_{B_2}$, the smoothing parameter for P-splines, and the proportion of variance explained ($pve$) for FPCR, were selected via 10-fold cross-validation, with details provided in the Supplementary Material. The trained models were then used to predict the logarithm of hurricane counts on the test set.

The performance of each method was evaluated by computing the root mean squared error (RMSE, $(\sum_{i=1}^{n}(\hat{y}_i - y_i)^2/n)^{1/2}$) and mean absolute error (MAE, $\sum_{i=1}^{n}|\hat{y}_i - y_i|/n$) on the test set, as summarized in Table 5. Predictions issued in June by CSU were also included for comparison. The CSU forecasts rely on four predictors: (i) ECMWF-predicted September SSTs in the equatorial Pacific (issued on May 1), (ii) April–May SSTs in the eastern North Atlantic, (iii) April–May 200-mb zonal winds in the tropical Pacific, and (iv) May sea level pressure in the central North Atlantic. Historical CSU forecasts are available at https://tropical.colostate.edu/archive.html. Despite relying solely on MAM-SSTs, the $\text{GDS}_{B_2}$ method achieves a 12.1% reduction in RMSE and an 11.6% reduction in MAE compared to CSU's June predictions.

|  | P-splines | FPCR | $\text{GDS}_{B_1}$ | $\text{GDS}_{B_2}$ | CSU |
|---|---|---|---|---|---|
| RMSE | 0.407 (0.068) | 0.415 (0.077) | 0.394 (0.069) | 0.402 (0.074) | 0.447 (0.077) |
| MAE | 0.316 (0.052) | 0.322 (0.058) | 0.308 (0.051) | 0.312 (0.055) | 0.352 (0.067) |

Table 5: Root mean squared errors (RMSEs) and mean absolute errors (MAEs) for predictions over 2005–2023. Standard errors (reported in parentheses) are obtained via a moving block bootstrap applied to the test-period forecast errors (block length = 5 years) to account for potential serial dependence across years. Results were qualitatively similar for block lengths between 3 and 5 years.

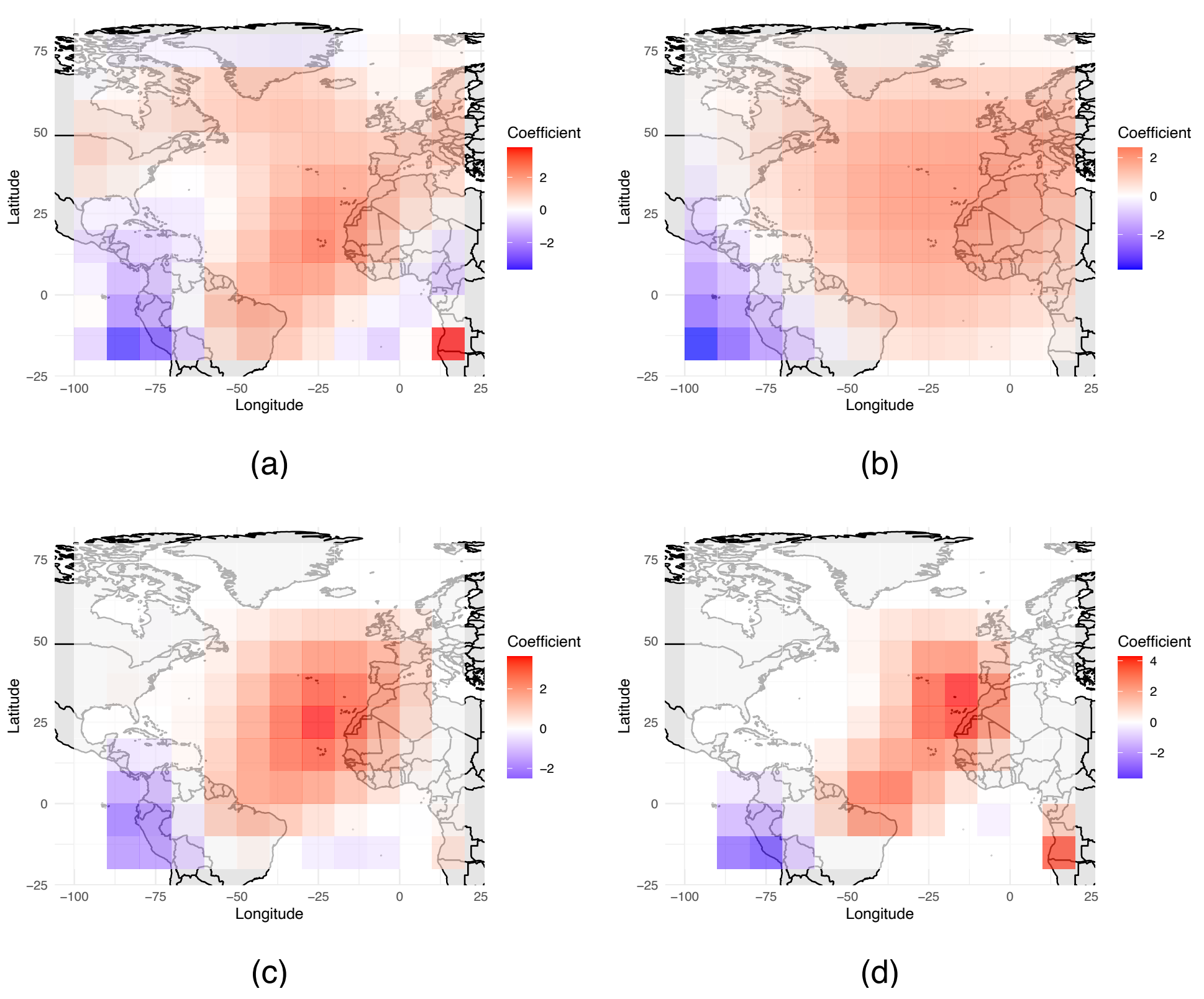

Figure 5: Estimated coefficient surfaces from (a) P-splines, (b) FPCR, (c) $\text{GDS}_{B_1}$, and (d) $\text{GDS}_{B_2}$.

Figure 5 shows that the estimated $\beta$ surface from proposed methods effectively captures the regions where correlations between MAM-SSTs and the logarithm of hurricane counts are significant at the 0.05 level. These regions closely align with findings in Bell and Chelliah (2006), Vimont and Kossin (2007), Saunders and Lea (2008), and Davis et al. (2015).

# 6 DISCUSSION

We have developed the Generalized Dantzig Selector (GDS) method for estimating the coefficient function in scalar-on-image regression models via a basis expansion framework. Additionally, we have provided a non-asymptotic theoretical analysis of our estimator, establishing finite-sample error bounds under mild conditions. Results from both simulation studies and real-world data applications demonstrate that our method effectively produces a smooth and interpretable coefficient surface while accurately recovering zero (non-predictive) and nonzero (predictive) regions.

While our current work focuses on models with a single image predictor, several directions for future research remain. One promising extension is to accommodate models involving multiple image-based predictors or multimodal data, where spatial information from different sources may jointly influence the response. Another important direction is to generalize the framework to broader classes of scalar-on-image models, such as those with non-Gaussian outcomes or generalized linear model (GLM) structures. These extensions would further enhance the flexibility and applicability of the GDS approach in modern high-dimensional imaging studies.

## ACKNOWLEDGMENTS

Funding was provided in part by R21AR083495-01A1, R01AR080742-02S1, and UA Eighteenth Mile TRIF Funding.

## DISCLOSURE STATEMENT

The authors report there are no competing interests to declare.

## SUPPLEMENTARY MATERIAL

**Proofs.** Proofs of all theoretical results are provided in the online supplementary material.

**Code and Data.** All code and curated data are available at https://github.com/Sja-88/GDS_Repo. The raw SST data are available from the NOAA, and the raw hurricane data can be accessed from the IBTrACS.

# Interpretable Scalar-on-Image Linear Regression Models via the Generalized Dantzig Selector

## SUPPLEMENTARY MATERIAL

# S1 PROOFS

Before presenting the proof of Theorem 1, we introduce several technical lemmas, namely Lemmas 1–6. Among these, Lemmas 1–3 also play a key role in the proof of Theorem 2. Lemmas 4 and 5 are slight modifications of Lemma 3.1 and related results in Candes and Tao (2007). Complete proofs for all lemmas are provided.

## S1.1 Lemma 1

Lemma 1 applies generally to the basis representation introduced in (2). It also holds under the orthogonal decomposition of $\beta$, where $\eta$ is replaced by $\eta^*$ and $\mathcal{E}$ by $\mathcal{E}_b$.

**Lemma 1.** *Suppose $\epsilon_i \sim \mathcal{N}(0, \sigma^2)$ independently for $i = 1, \ldots, n$, with $\sigma > 0$, and assume there exists a constant $M < \infty$ such that $\|x_i\|_{L_2(\Omega)} \leq M$ for all $i$. Let $\eta$ denote the coefficient vector under the selected basis $b$ as defined in (2), and let $\mathcal{E} = \|e\|_{L_2(\Omega)}$. Set*

$$\lambda = C\sigma\sqrt{\frac{\log p}{n}} + M\mathcal{E},$$

*for some constant $C > \sqrt{2}$. Then, with probability at least $1 - p^{1-C^2/2}$,*

$$\left\|\frac{1}{n}N^{-1}X^\top(y - X\eta)\right\|_{\ell_\infty} \leq \lambda.$$

*Proof of Lemma 1.* We aim to determine the value of $\lambda$ such that $\left\|\frac{1}{n}N^{-1}X^\top(y - X\eta)\right\|_{\ell_\infty} \leq \lambda$ holds with high probability. Define

$$v_i = \int_0^1 \int_0^1 x_i(t, s)e(t, s)\, dtds, \tag{S1}$$

for $i = 1, \ldots, n$, and let $v = (v_1, \ldots, v_n)^\top$. Then we have

$$\begin{aligned}\frac{1}{n} N^{-1} X^\top (y - X\eta) &= \frac{1}{n} N^{-1} X^\top \epsilon^* \\ &= \frac{1}{n} N^{-1} X^\top \epsilon + \frac{1}{n} N^{-1} X^\top v.\end{aligned}$$

The $j$th element of $\frac{1}{n} N^{-1} X^\top (y - X\eta)$ is given by

$$\left(\frac{1}{n} N^{-1} X^\top (y - X\eta)\right)_j = \frac{1}{n} N_j^{-1} X_j^\top \epsilon + \frac{1}{n} N_j^{-1} X_j^\top v.$$

Note that $N_j = \sqrt{\frac{1}{n}\sum_{i=1}^n X_{ij}^2}$. The expression above can be rewritten as

$$\left(\frac{1}{n} N^{-1} X^\top (y - X\eta)\right)_j = \frac{1}{\sqrt{n}} \tilde{X}_j^\top \epsilon + \frac{1}{\sqrt{n}} \tilde{X}_j^\top v, \tag{S2}$$

where $\tilde{X}_j = X_j / \sqrt{\sum_{i=1}^n X_{ij}^2}$, for $j = 1, \ldots, p$. Notice that $\|\tilde{X}_j\|_{\ell_2} = 1$. It follows that

$$\tilde{X}_j^\top \epsilon \sim \mathcal{N}(0, \sigma^2)$$

and

$$|\tilde{X}_j^\top v| \leq \|\tilde{X}_j\|_{\ell_2} \|v\|_{\ell_2} = \|v\|_{\ell_2}.$$

Now we bound the second term in (S2). Since $|v_i| \leq \|x_i\|_{L_2(\Omega)} \|e\|_{L_2(\Omega)} \leq M\mathcal{E}$ for each $i = 1, \ldots, n$, and $\|v\|_{\ell_2} \leq \sqrt{n}\|v\|_{\ell_1}$, we have

$$\left|\frac{1}{\sqrt{n}} \tilde{X}_j^\top v\right| \leq \|v\|_{\ell_1} \leq M\mathcal{E}, \quad \text{for all } j = 1, \ldots, p.$$

To bound the first term in (S2), note that for $Z \sim \mathcal{N}(0, \sigma^2)$ and any $t \geq 0$, we have

$$\begin{aligned}P(Z \geq t) \exp\left\{\frac{t^2}{2\sigma^2}\right\} &= \int_t^\infty \frac{1}{\sqrt{2\pi}\sigma} \exp\left\{-\frac{x^2}{2\sigma^2}\right\} dx \cdot \exp\left\{\frac{t^2}{2\sigma^2}\right\} \\ &= \int_t^\infty \frac{1}{\sqrt{2\pi}\sigma} \exp\left\{-\frac{(x-t)^2}{2\sigma^2}\right\} \exp\left\{\frac{2t^2 - 2xt}{2\sigma^2}\right\} dx \\ &\leq \int_t^\infty \frac{1}{\sqrt{2\pi}\sigma} \exp\left\{-\frac{(x-t)^2}{2\sigma^2}\right\} dx \\ &\leq \frac{1}{2}.\end{aligned}$$

Therefore, $P(|Z| \geq t) \leq 2P(Z \geq t) \leq \exp\left\{-\frac{t^2}{2\sigma^2}\right\}$ for any $t \geq 0$.

Let $Z_j = \tilde{X}_j^\top \epsilon$, then $Z_j \sim \mathcal{N}(0, \sigma^2)$ for $j = 1, \ldots, p$. Furthermore,

$$\begin{aligned} P\left(\sup_j |Z_j| \geq t\right) &= P\left(\bigcup_{j=1}^{p} \{|Z_j| \geq t\}\right) \\ &\leq \sum_{j=1}^{p} P(|Z_j| \geq t) \\ &\leq p \exp\left\{-\frac{t^2}{2\sigma^2}\right\}. \end{aligned}$$

Setting $t = C\sigma\sqrt{\log p}$ gives

$$P\left(\sup_j |Z_j| \geq C\sigma\sqrt{\log p}\right) \leq p \exp\left\{-\frac{C^2 \log p}{2}\right\} = p^{1-C^2/2}.$$

Therefore, for any $C > \sqrt{2}$, with probability at least $1 - p^{1-C^2/2}$, we have

$$\sup_j \left|\frac{1}{\sqrt{n}} \tilde{X}_j^\top \epsilon\right| \leq C\sigma\sqrt{\frac{\log p}{n}}.$$

Combining the above with (S2), we conclude that

$$\begin{aligned} \left\|\frac{1}{n} N^{-1} X^\top (y - X\eta)\right\|_{\ell_\infty} &\leq \sup_j \left|\frac{1}{\sqrt{n}} \tilde{X}_j^\top \epsilon\right| + \sup_j \left|\frac{1}{\sqrt{n}} \tilde{X}_j^\top v\right| \\ &\leq C\sigma\sqrt{\frac{\log p}{n}} + M\mathcal{E}, \end{aligned}$$

with probability at least $1 - p^{1-C^2/2}$ for any $C > \sqrt{2}$. This completes the proof. □

## S1.2 Lemma 2

**Lemma 2.** *Let $\gamma = A\eta$. Suppose $\gamma$ has at most $S$ nonzero components, and let $\mathcal{T}_0 = \{j : \gamma_j \neq 0\}$ denote its true support. If*

$$\left\|\frac{1}{n} N^{-1} X^\top (y - X\eta)\right\|_{\ell_\infty} \leq \lambda,$$

*then for $\hat{\eta}$ obtained as the solution to (6), defining $h = \hat{\gamma} - \gamma$ with $\hat{\gamma} = A\hat{\eta}$ and letting $\mathcal{T}_1$ be the $S$ largest entries of $h$ outside $\mathcal{T}_0$, we have*

$$\|h_{\mathcal{T}_{01}^\complement}\|_{\ell_2} \leq \frac{1}{\sqrt{S}} \|h_{\mathcal{T}_0^\complement}\|_{\ell_1}, \tag{S3}$$

*where* $\mathcal{T}_{01} = \mathcal{T}_0 \cup \mathcal{T}_1$, *and*

$$\|h_{\mathcal{T}_0^{\complement}}\|_{\ell_1} \leq \|h_{\mathcal{T}_0}\|_{\ell_1}. \tag{S4}$$

*Proof of Lemma 2.* Following the argument in Candes and Tao (2007), we first observe that the $k$th largest entry of $h_{\mathcal{T}_0^{\complement}}$ satisfies

$$k \cdot |h_{\mathcal{T}_0^{\complement}}|_{(k)} \leq \|h_{\mathcal{T}_0^{\complement}}\|_{\ell_1},$$

which implies

$$\|h_{\mathcal{T}_{01}^{\complement}}\|_{\ell_2}^2 = \sum_{k \geq S+1} |h_{\mathcal{T}_0^{\complement}}|_{(k)}^2 \leq \|h_{\mathcal{T}_0^{\complement}}\|_{\ell_1}^2 \sum_{k \geq S+1} \frac{1}{k^2} \leq \frac{1}{S} \|h_{\mathcal{T}_0^{\complement}}\|_{\ell_1}^2,$$

establishing (S3).

Next, since both $\hat{\eta}$ and $\eta$ are feasible solutions to problem (6), it follows that

$$\|\hat{\gamma}\|_{\ell_1} \leq \|\gamma\|_{\ell_1}.$$

As shown similarly in Candes and Tao (2007) and Donoho and Huo (2001), this implies

$$\|h_{\mathcal{T}_0^{\complement}}\|_{\ell_1} \leq \|h_{\mathcal{T}_0}\|_{\ell_1}.$$

To prove this, decompose $\hat{\gamma} = \gamma + h$, and note that

$$\|\gamma + h\|_{\ell_1} - \|\gamma\|_{\ell_1} = \|h_{\mathcal{T}_0^{\complement}}\|_{\ell_1} + \|(\gamma + h)_{\mathcal{T}_0}\|_{\ell_1} - \|\gamma_{\mathcal{T}_0}\|_{\ell_1} \leq 0.$$

Rewriting the above yields

$$\begin{aligned} \|h_{\mathcal{T}_0^{\complement}}\|_{\ell_1} + \|\gamma_{\mathcal{T}_0}\|_{\ell_1} - \|h_{\mathcal{T}_0}\|_{\ell_1} &\leq \|h_{\mathcal{T}_0^{\complement}}\|_{\ell_1} + \|(\gamma + h)_{\mathcal{T}_0}\|_{\ell_1} \\ &\leq \|\gamma_{\mathcal{T}_0}\|_{\ell_1}, \end{aligned}$$

which implies

$$\|h_{\mathcal{T}_0^{\complement}}\|_{\ell_1} \leq \|h_{\mathcal{T}_0}\|_{\ell_1}.$$

This concludes the proof of (S4). $\square$

## S1.3 Lemma 3

**Lemma 3.** *Let $V = XA^+$ and $\gamma^* = A\eta^*$. The model (6) is equivalent to the following optimization problem*

$$\min_{\gamma \in \mathbb{R}^L} \|\gamma\|_{\ell_1}, \quad \textit{subject to} \quad \left\| \frac{1}{n} N^{-1} X^\top (y - V\gamma) \right\|_{\ell_\infty} \leq \lambda.$$

*Let $\hat{\gamma}$ be the solution to it and suppose that $\gamma^*$ is also a feasible solution. Then the difference $h = \hat{\gamma} - \gamma^*$ satisfies*

$$\frac{1}{n} \|V^\top V h\|_{\ell_\infty} \leq \frac{\sqrt{L} N_{\max}}{\sigma_{\min}(A)} \left\| \frac{1}{n} N^{-1} X^\top V h \right\|_{\ell_\infty}. \tag{S5}$$

*Proof of Lemma 3.* By construction, the matrix $A \in \mathbb{R}^{L \times p}$, where $L \geq p$, and there exists a pseudo-inverse $A^+ \in \mathbb{R}^{p \times L}$ such that $A^+ A = I_p$. Recall that $y = V\gamma + \epsilon^*$, where $V = XA^+$ and $\gamma = A\eta$. Solving the optimization problem defined in (6) is equivalent to obtaining $\hat{\gamma}$ by solving

$$\min_{\gamma \in \mathbb{R}^L} \|\gamma\|_{\ell_1}, \quad \text{subject to} \quad \left\| \frac{1}{n} N^{-1} X^\top (y - V\gamma) \right\|_{\ell_\infty} \leq \lambda.$$

Let $h = \hat{\gamma} - \gamma^*$. Since both $\hat{\gamma}$ and $\gamma^*$ are feasible, it follows from the triangle inequality that

$$\left\| \frac{1}{n} N^{-1} X^\top V h \right\|_{\ell_\infty} \leq 2\lambda. \tag{S6}$$

Let $\mathcal{T}_0 = \{j : \gamma_j^* \neq 0\}$ be the support of $\gamma^*$, with $|\mathcal{T}_0| \leq S$. Let $\mathcal{T}_1$ be the indices of the $S$ largest entries of $h$ outside $\mathcal{T}_0$, and define $\mathcal{T}_{01} = \mathcal{T}_0 \cup \mathcal{T}_1$. From Lemma 2, we know

$$\|h_{\mathcal{T}_0^\complement}\|_{\ell_1} \leq \|h_{\mathcal{T}_0}\|_{\ell_1}.$$

Next, we have

$$\|V^\top V h\|_{\ell_\infty} = \|{A^+}^\top X^\top V h\|_{\ell_\infty} \leq \|N A^+\|_1 \left\| N^{-1} X^\top V h \right\|_{\ell_\infty},$$

where the inequality follows from $\|Mv\|_{\ell_\infty} \leq \|M\|_\infty \|v\|_{\ell_\infty}$ and $\|M^\top\|_\infty = \|M\|_1$. Here, $\|\cdot\|_\infty$, $\|\cdot\|_1$, and $\|\cdot\|_2$ denote matrix norms, defined as

$$\|M\|_\infty = \max_{1 \leq i \leq q} \sum_{j=1}^{r} |m_{ij}|, \qquad \|M\|_1 = \max_{1 \leq j \leq r} \sum_{i=1}^{q} |m_{ij}|, \quad \text{and} \quad \|M\|_2 = \sqrt{\sigma_{max}(M^\top M)},$$

for a matrix $M = (m_{ij}) \in \mathbb{R}^{q\times r}$.

Moreover,

$$\|NA^+\|_1 \le N_{\max}\|A^+\|_1, \quad \text{and} \quad \|A^+\|_1 \le \sqrt{L}\|A^+\|_2 = \frac{\sqrt{L}}{\sigma_{\min}(A)},$$

where the last equality uses the singular value decomposition of $A$. Therefore,

$$\frac{1}{n}\|V^\top Vh\|_{\ell_\infty} \le \frac{\sqrt{L}N_{\max}}{\sigma_{\min}(A)}\left\|\frac{1}{n}N^{-1}X^\top Vh\right\|_{\ell_\infty},$$

which establishes (S5). $\square$

## S1.4 Lemma 4

**Lemma 4** (cf. Lemma 3.1 in Candes and Tao (2007))**.** *Let $\mathcal{T}_0$ be an index set with $|\mathcal{T}_0| = S$, where $3S \le L$. Suppose $V \in \mathbb{R}^{n\times L}$ satisfies $\phi_{2S} > \theta_{S,2S}$. For any $h \in \mathbb{R}^L$, let $\mathcal{T}_1$ denote the set of the $S$ largest entries of $h$ outside of $\mathcal{T}_0$, and set $\mathcal{T}_{01} = \mathcal{T}_0 \cup \mathcal{T}_1$. Then*

$$\|h_{\mathcal{T}_{01}}\|_{\ell_2} \le \frac{1}{n\phi_{2S}}\|V_{\mathcal{T}_{01}}^\top Vh\|_{\ell_2} + \frac{\theta_{S,2S}}{\phi_{2S}S^{1/2}}\|h_{\mathcal{T}_0^\complement}\|_{\ell_1}.$$

*Proof of Lemma 4.* Consider the restricted transformation

$$V_{\mathcal{T}_{01}} : \mathbb{R}^{|\mathcal{T}_{01}|} \to \mathbb{R}^n, \quad V_{\mathcal{T}_{01}}c_{\mathcal{T}_{01}} := \sum_{j\in\mathcal{T}_{01}} c_j V^j.$$

Let $\mathcal{W} \subset \mathbb{R}^n$ denote the span of $\{V^j : j \in \mathcal{T}_{01}\}$. Then $\mathcal{W}$ is the range of $V_{\mathcal{T}_{01}}$ and the orthogonal complement of the kernel of $V_{\mathcal{T}_{01}}^\top$. Thus, $\mathbb{R}^n = \mathcal{W} \oplus \mathcal{W}^\perp$.

Since $\phi_{2S} > 0$, we know that $V_{\mathcal{T}_{01}}$ is a bijection from $\mathbb{R}^{|\mathcal{T}_{01}|}$ to $\mathcal{W}$, with singular values greater than or equal to $\sqrt{n\phi_{2S}}$. By Definition 1, for any $c \in \mathbb{R}^L$, we have

$$\sqrt{\phi_{2S}}\|c_{\mathcal{T}_{01}}\|_{\ell_2} \le \|V_{\mathcal{T}_{01}}c_{\mathcal{T}_{01}}\|_{\ell_2}/\sqrt{n}.$$

Let $P_{\mathcal{W}}$ denote the orthogonal projection onto $\mathcal{W}$. For each $v \in \mathbb{R}^n$, we have $V_{\mathcal{T}_{01}}^\top v = V_{\mathcal{T}_{01}}^\top P_{\mathcal{W}}v$. Since $V_{\mathcal{T}_{01}}^\top$ and $V_{\mathcal{T}_{01}}$ share the same singular values, it follows that

$$\sqrt{\phi_{2S}}\|P_{\mathcal{W}}v\|_{\ell_2} \le \|V_{\mathcal{T}_{01}}^\top v\|_{\ell_2}/\sqrt{n}.$$

Applying this to $v := Vh$ gives

$$\|P_{\mathcal{W}}Vh\|_{\ell_2} \leq (\phi_{2S})^{-1/2}\|V_{\mathcal{T}_{01}}^{\top}Vh\|_{\ell_2}/\sqrt{n}. \tag{S7}$$

We now divide $\mathcal{T}_0^{\complement}$ into subsets of size $S$, where $\mathcal{T}_1$ contains the indices of the $S$ largest coefficients of $h_{\mathcal{T}_0^{\complement}}$, $\mathcal{T}_2$ contains the indices of the next $S$ largest coefficients, and so on. Decompose $P_{\mathcal{W}}Vh$ as

$$P_{\mathcal{W}}Vh = P_{\mathcal{W}}Vh_{\mathcal{T}_{01}} + \sum_{j\geq 2} P_{\mathcal{W}}Vh_{\mathcal{T}_j}.$$

By definition of $\mathcal{W}$, we have $Vh_{\mathcal{T}_{01}} \in \mathcal{W}$ and $P_{\mathcal{W}}Vh_{\mathcal{T}_{01}} = Vh_{\mathcal{T}_{01}}$.

Since $P_{\mathcal{W}}$ projects onto $\mathcal{W}$, there exists some $c \in \mathbb{R}^L$ such that $P_{\mathcal{W}}Vh_{\mathcal{T}_j} = \sum_{j\in\mathcal{T}_{01}} c_j V^j = Vc_{\mathcal{T}_{01}}$, and

$$\|P_{\mathcal{W}}Vh_{\mathcal{T}_j}\|_{\ell_2}^2 = \langle P_{\mathcal{W}}Vh_{\mathcal{T}_j}, Vh_{\mathcal{T}_j}\rangle = \langle Vc_{\mathcal{T}_{01}}, Vh_{\mathcal{T}_j}\rangle.$$

By using the restricted orthogonality constant $\theta_{S,2S}$ defined in Definition 2, we have

$$\begin{aligned}
\langle Vc_{\mathcal{T}_{01}}, Vh_{\mathcal{T}_j}\rangle &\leq n\theta_{S,2S}\|c_{\mathcal{T}_{01}}\|_{\ell_2}\|h_{\mathcal{T}_j}\|_{\ell_2} \\
&\leq \frac{\sqrt{n}\theta_{S,2S}}{\sqrt{\phi_{2S}}}\|Vc_{\mathcal{T}_{01}}\|_{\ell_2}\|h_{\mathcal{T}_j}\|_{\ell_2} \\
&= \frac{\sqrt{n}\theta_{S,2S}}{\sqrt{\phi_{2S}}}\|P_{\mathcal{W}}Vh_{\mathcal{T}_j}\|_{\ell_2}\|h_{\mathcal{T}_j}\|_{\ell_2}.
\end{aligned}$$

Thus,

$$\|P_{\mathcal{W}}Vh_{\mathcal{T}_j}\|_{\ell_2} \leq \frac{\sqrt{n}\theta_{S,2S}}{\sqrt{\phi_{2S}}}\|h_{\mathcal{T}_j}\|_{\ell_2}. \tag{S8}$$

We now develop an upper bound for $\sum_{j\geq 2}\|h_{\mathcal{T}_j}\|_{\ell_2}$ following Candes et al. (2005) and Candes and Tao (2007). By construction, for each component $h_{\mathcal{T}_{j+1}}[i]$,

$$|h_{\mathcal{T}_{j+1}}[i]| \leq \|h_{\mathcal{T}_j}\|_{\ell_1}/S,$$

which implies

$$\|h_{\mathcal{T}_{j+1}}\|_{\ell_2}^2 \leq \|h_{\mathcal{T}_j}\|_{\ell_1}^2/S,$$

and hence,

$$\sum_{j\geq 2}\|h_{\mathcal{T}_j}\|_{\ell_2} \leq S^{-1/2}\sum_{j\geq 1}\|h_{\mathcal{T}_j}\|_{\ell_1} = S^{-1/2}\|h_{\mathcal{T}_0^{\complement}}\|_{\ell_1}.$$

In summary,

$$\begin{aligned}\|P_{\mathcal{W}}Vh\|_{\ell_2} &\geq \|P_{\mathcal{W}}Vh_{\mathcal{T}_{01}}\|_{\ell_2} - \sum_{j\geq 2}\|P_{\mathcal{W}}Vh_{\mathcal{T}_j}\|_{\ell_2} \\ &\geq \|Vh_{\mathcal{T}_{01}}\|_{\ell_2} - \frac{\sqrt{n}\theta_{S,2S}}{\sqrt{\phi_{2S}}}S^{-1/2}\|h_{\mathcal{T}_0^{\complement}}\|_{\ell_1} \\ &\geq \sqrt{n\phi_{2S}}\|h_{\mathcal{T}_{01}}\|_{\ell_2} - \frac{\sqrt{n}\theta_{S,2S}}{\sqrt{\phi_{2S}}}S^{-1/2}\|h_{\mathcal{T}_0^{\complement}}\|_{\ell_1}.\end{aligned}$$

Combining this with (S7) gives

$$\|h_{\mathcal{T}_{01}}\|_{\ell_2} \leq \frac{1}{n\phi_{2S}}\|V_{\mathcal{T}_{01}}^{\top}Vh\|_{\ell_2} + \frac{\theta_{S,2S}}{\phi_{2S}S^{1/2}}\|h_{\mathcal{T}_0^{\complement}}\|_{\ell_1},$$

which completes the proof. □

## S1.5 Lemma 5

**Lemma 5.** *Let $\gamma = A\eta$, where $\eta$ is the coefficient vector under the selected basis $b$. Suppose $\gamma$ has at most $S$ nonzero components, and let $\mathcal{T}_0 = \{j : \gamma_j \neq 0\}$ denote its true support. Define $V = XA^{+}$, and assume $V$ satisfies $\phi_{2S} > \theta_{S,2S}$. If*

$$\left\|\frac{1}{n}N^{-1}X^{\top}(y - X\eta)\right\|_{\ell_\infty} \leq \lambda,$$

*then for $\hat{\eta}$ obtained as the solution to (6), defining $h = \hat{\gamma} - \gamma$ with $\hat{\gamma} = A\hat{\eta}$ and letting $\mathcal{T}_1$ be the $S$ largest entries of $h$ outside $\mathcal{T}_0$, we have*

$$\|h\|_{\ell_2} \leq \frac{\sqrt{2}}{n(\phi_{2S} - \theta_{S,2S})}\|V_{\mathcal{T}_{01}}^{\top}Vh\|_{\ell_2},$$

*where $\mathcal{T}_{01} = \mathcal{T}_0 \cup \mathcal{T}_1$.*

*Proof of Lemma 5.* Since $\|h\|_{\ell_2}^2 = \|h_{\mathcal{T}_{01}}\|_{\ell_2}^2 + \|h_{\mathcal{T}_{01}^{\complement}}\|_{\ell_2}^2$, combining this with inequality (S3), we obtain

$$\|h\|_{\ell_2}^2 \leq \|h_{\mathcal{T}_{01}}\|_{\ell_2}^2 + \frac{1}{S}\|h_{\mathcal{T}_0^{\complement}}\|_{\ell_1}^2. \tag{S9}$$

From Lemma 4, we have

$$\|h_{\mathcal{T}_{01}}\|_{\ell_2} \leq \frac{1}{n\phi_{2S}}\|V_{\mathcal{T}_{01}}^{\top} V h\|_{\ell_2} + \frac{\theta_{S,2S}}{\phi_{2S}\sqrt{S}}\|h_{\mathcal{T}_0^{\complement}}\|_{\ell_1}.$$

In addition, we observe that

$$\|h_{\mathcal{T}_0}\|_{\ell_1} \leq \sqrt{S}\|h_{\mathcal{T}_0}\|_{\ell_2} \leq \sqrt{S}\|h_{\mathcal{T}_{01}}\|_{\ell_2}.$$

Substituting into the previous bound and using inequality (S4), we get

$$\|h_{\mathcal{T}_{01}}\|_{\ell_2} \leq \frac{1}{n\phi_{2S}}\|V_{\mathcal{T}_{01}}^{\top} V h\|_{\ell_2} + \frac{\theta_{S,2S}}{\phi_{2S}}\|h_{\mathcal{T}_{01}}\|_{\ell_2}.$$

Rearranging terms gives

$$\left(1 - \frac{\theta_{S,2S}}{\phi_{2S}}\right)\|h_{\mathcal{T}_{01}}\|_{\ell_2} \leq \frac{1}{n\phi_{2S}}\|V_{\mathcal{T}_{01}}^{\top} V h\|_{\ell_2},$$

which implies

$$\|h_{\mathcal{T}_{01}}\|_{\ell_2} \leq \frac{1}{n(\phi_{2S} - \theta_{S,2S})}\|V_{\mathcal{T}_{01}}^{\top} V h\|_{\ell_2}.$$

Now, combining (S9) with (S4), we have

$$\begin{aligned}
\|h\|_{\ell_2}^2 &\leq \|h_{\mathcal{T}_{01}}\|_{\ell_2}^2 + \frac{1}{S}\|h_{\mathcal{T}_0^{\complement}}\|_{\ell_1}^2 \\
&\leq \|h_{\mathcal{T}_{01}}\|_{\ell_2}^2 + \frac{1}{S}\|h_{\mathcal{T}_0}\|_{\ell_1}^2 \\
&\leq 2\|h_{\mathcal{T}_{01}}\|_{\ell_2}^2,
\end{aligned}$$

which yields

$$\|h\|_{\ell_2} \leq \frac{\sqrt{2}}{n(\phi_{2S} - \theta_{S,2S})}\|V_{\mathcal{T}_{01}}^{\top} V h\|_{\ell_2}.$$

This completes the proof. □

## S1.6 Lemma 6

We now focus on the orthogonal decomposition. The following lemma bounds the difference between $\hat{\eta}$ and $\eta^*$:

**Lemma 6.** *Suppose $\gamma^* = A\eta^*$ has at most $S$ nonzero components and that $\phi_{2S} > \theta_{S,2S}$ holds for $V = XA^{+}$. Let $\hat{\gamma} = A\hat{\eta}$, where $\hat{\eta}$ is obtained by solving (6). If*

$$\left\| \frac{1}{n} N^{-1} X^\top (y - X\eta^*) \right\|_{\ell_\infty} \le \lambda,$$

*then*

$$\|\hat{\gamma} - \gamma^*\|_{\ell_2} \le \frac{4\sqrt{L} N_{\max}}{(\phi_{2S} - \theta_{S,2S})\, \sigma_{\min}(A)} \lambda \sqrt{S}, \tag{S10}$$

*and*

$$\|\hat{\eta} - \eta^*\|_{\ell_2} \le \frac{4\sqrt{L} N_{\max}}{(\phi_{2S} - \theta_{S,2S})\, \sigma_{\min}^2(A)} \lambda \sqrt{S}, \tag{S11}$$

*where $N_{\max} = \max_{j=1,\ldots,p} N_j$ and $\sigma_{\min}(A)$ is the smallest singular value of $A$.*

*Proof of Lemma 6.* Let $h = \hat{\gamma} - \gamma^*$, and let $\mathcal{T}_0$ denote the support of $\gamma^*$, i.e.,

$$\mathcal{T}_0 = \{1 \le j \le L : \gamma_j^* \ne 0\},$$

so that $|\mathcal{T}_0| \le S$. Let $\mathcal{T}_1$ be the index set of the $S$ largest entries of $h$ outside $\mathcal{T}_0$, and define $\mathcal{T}_{01} = \mathcal{T}_0 \cup \mathcal{T}_1$.

From Lemma 5, we have

$$\|h\|_{\ell_2} \le \frac{\sqrt{2}}{n(\phi_{2S} - \theta_{S,2S})} \|V_{\mathcal{T}_{01}}^\top V h\|_{\ell_2}. \tag{S12}$$

To bound $\|V_{\mathcal{T}_{01}}^\top V h\|_{\ell_2}$, note that

$$\|V_{\mathcal{T}_{01}}^\top V h\|_{\ell_2} \le \sqrt{2S} \cdot \|V^\top V h\|_{\ell_\infty}.$$

Combining this with (S12), inequality (S5), and bound (S6), we obtain

$$\begin{aligned}
\|h\|_{\ell_2} &\le \frac{\sqrt{2}}{n(\phi_{2S} - \theta_{S,2S})} \cdot \sqrt{2S} \cdot \|V^\top V h\|_{\ell_\infty} \\
&\le \frac{2\sqrt{L} N_{\max}}{(\phi_{2S} - \theta_{S,2S})\sigma_{\min}(A)} \sqrt{S} \left\| \frac{1}{n} N^{-1} X^\top V h \right\|_{\ell_\infty} \\
&\le \frac{4\sqrt{L} N_{\max}}{(\phi_{2S} - \theta_{S,2S})\sigma_{\min}(A)} \lambda \sqrt{S}.
\end{aligned}$$

Finally, since $\hat{\eta} - \eta^* = A^+ h$, we have

$$\|\hat{\eta} - \eta^*\|_{\ell_2} = \|A^+ h\|_{\ell_2} \leq \frac{\|h\|_{\ell_2}}{\sigma_{\min}(A)},$$

which establishes the bound in (S11). $\square$

## S1.7 Proof of Theorem 1

*Proof of Theorem 1.* From Lemmas 1 and 6, we know that inequalities (S10) and (S11) hold simultaneously with probability at least $1 - p^{1-C^2/2}$, for some constant $C > \sqrt{2}$ and

$$\lambda = C\sigma\sqrt{\frac{\log p}{n}} + M\mathcal{E}_b.$$

By the definition of the $L_2(\Omega)$ norm, the triangle inequality, and the Cauchy–Schwarz inequality, we obtain

$$\begin{aligned}
\|\hat{\beta} - \beta\|_{L_2(\Omega)} &\leq \left\|\sum_{j=1}^{p}(\hat{\eta}_j - \eta_j^*)b_j(t,s)\right\|_{L_2(\Omega)} + \quad \|e\|_{L_2(\Omega)} \\
&\leq \sqrt{\sum_{j=1}^{p}\left(\iint_{\Omega} b_j^2(t,s)\,dtds\right)} \cdot \|\hat{\eta} - \eta^*\|_{\ell_2} + \mathcal{E}_b. \qquad \text{(S13)}
\end{aligned}$$

Inequality (9) then follows directly from (S13) and (S11).

To bound the prediction error $\|\hat{\mu} - \mu\|_n$, let $\mu = X\eta^* + v$, where $v$ is defined in (S1), and $\hat{\mu} = X\hat{\eta}$. By the triangle inequality,

$$\|\hat{\mu} - \mu\|_n \leq \frac{1}{\sqrt{n}}\|X(\hat{\eta} - \eta^*)\|_{\ell_2} + \|v\|_n,$$

where, from Lemma 1,

$$\|v\|_n = \frac{1}{\sqrt{n}}\|v\|_{\ell_2} \leq \|v\|_{\ell_1} \leq M\mathcal{E}_b.$$

To bound the first term, note that from Lemma 3, we have

$$\frac{1}{n}\|V^\top V h\|_{\ell_\infty} \leq \frac{\sqrt{L}N_{\max}}{\sigma_{\min}(A)}\left\|\frac{1}{n}N^{-1}X^\top V h\right\|_{\ell_\infty},$$

and

$$\left\|\frac{1}{n}N^{-1}X^\top Vh\right\|_{\ell_\infty} \leq 2\lambda.$$

Combining these with (S10), we get

$$\begin{aligned}
\frac{1}{n}\|X(\hat{\eta}-\eta^*)\|_{\ell_2}^2 &= \frac{1}{n}h^\top V^\top Vh \\
&\leq \frac{1}{n}\|V^\top Vh\|_{\ell_\infty}\cdot\|h\|_{\ell_1} \\
&\leq \frac{\sqrt{L}N_{\max}}{\sigma_{\min}(A)}\cdot\left\|\frac{1}{n}N^{-1}X^\top Vh\right\|_{\ell_\infty}\cdot 2\|h_{\mathcal{T}_0}\|_{\ell_1} \\
&\leq \frac{4\sqrt{L}N_{\max}}{\sigma_{\min}(A)}\lambda\sqrt{S}\cdot\|h\|_{\ell_2} \\
&\leq \frac{16LN_{\max}^2}{(\phi_{2S}-\theta_{S,2S})\sigma_{\min}^2(A)}\lambda^2 S.
\end{aligned}$$

Taking square roots yields

$$\frac{1}{\sqrt{n}}\|X(\hat{\eta}-\eta^*)\|_{\ell_2} \leq \frac{4\sqrt{L}N_{\max}}{\sqrt{\phi_{2S}-\theta_{S,2S}}\,\sigma_{\min}(A)}\lambda\sqrt{S}.$$

Therefore, the prediction error is bounded as

$$\|\hat{\mu}-\mu\|_n \leq \frac{4\sqrt{L}N_{\max}}{\sqrt{\phi_{2S}-\theta_{S,2S}}\,\sigma_{\min}(A)}\lambda\sqrt{S} + M\mathcal{E}_b,$$

which establishes inequality (10) and completes the proof. □

Before presenting the proof of Theorem 2, we first introduce Lemma 7 and 8. These results serve as key building blocks for the proof of Theorem 2.

## S1.8 Lemma 7

**Lemma 7.** *Suppose $\gamma^* = A\eta^*$ has at most $S$ nonzero components and that $\kappa_1(S) > 0$ holds for $V = XA^+$. Let $\hat{\gamma} = A\hat{\eta}$, where $\hat{\eta}$ is obtained by solving (6). If*

$$\left\|\frac{1}{n}N^{-1}X^\top(y-X\eta^*)\right\|_{\ell_\infty} \leq \lambda,$$

*then*

$$\|\hat{\gamma} - \gamma^*\|_{\ell_1} \leq \frac{8\sqrt{L}N_{\max}}{\kappa_1^2(S)\,\sigma_{\min}(A)}\,\lambda S, \tag{S14}$$

*and*

$$\frac{1}{n}\|V(\hat{\gamma} - \gamma^*)\|_{\ell_2}^2 \leq \frac{16LN_{\max}^2}{\kappa_1^2(S)\,\sigma_{\min}^2(A)}\,\lambda^2 S, \tag{S15}$$

*where* $N_{\max} = \max_{j=1,\dots,p} N_j$ *and* $\sigma_{\min}(A)$ *is the smallest singular value of* $A$*.*

*Proof of Lemma 7.* By using (S5) and (S6), we bound

$$\begin{aligned}
\frac{1}{n}\|Vh\|_{\ell_2}^2 &= \frac{1}{n}h^\top V^\top V h \\
&\leq \frac{1}{n}\|V^\top V h\|_{\ell_\infty}\|h\|_{\ell_1} \\
&\leq \frac{\sqrt{L}N_{\max}}{\sigma_{\min}(A)}\left\|\frac{1}{n}N^{-1}X^\top V h\right\|_{\ell_\infty}\|h\|_{\ell_1} \\
&\leq \frac{2\sqrt{L}N_{\max}}{\sigma_{\min}(A)}\lambda\|h\|_{\ell_1} \\
&\leq \frac{2\sqrt{L}N_{\max}}{\sigma_{\min}(A)}\lambda\left(\|h_{\mathcal{T}_0}\|_{\ell_1} + \|h_{\mathcal{T}_0^\complement}\|_{\ell_1}\right) \\
&\leq \frac{4\sqrt{L}N_{\max}}{\sigma_{\min}(A)}\lambda\|h_{\mathcal{T}_0}\|_{\ell_1}.
\end{aligned}$$

Since $\|h_{\mathcal{T}_0}\|_{\ell_1} \leq \sqrt{S}\|h_{\mathcal{T}_0}\|_{\ell_2}$, we obtain

$$\frac{1}{n}\|Vh\|_{\ell_2}^2 \leq \frac{4\sqrt{L}N_{\max}}{\sigma_{\min}(A)}\lambda\sqrt{S}\|h_{\mathcal{T}_0}\|_{\ell_2}. \tag{S16}$$

Since $\kappa_1(S) > 0$, by Definition 3 (restricted eigenvalue condition), we also have

$$\frac{1}{n}\|Vh\|_{\ell_2}^2 \geq \kappa_1^2(S)\|h_{\mathcal{T}_0}\|_{\ell_2}^2.$$

Combining this with (S16) yields

$$\|h_{\mathcal{T}_0}\|_{\ell_2} \leq \frac{4\sqrt{L}N_{\max}}{\kappa_1^2(S)\sigma_{\min}(A)}\lambda\sqrt{S}. \tag{S17}$$

Therefore,

$$\begin{aligned}\|h\|_{\ell_1} &= \|h_{\mathcal{T}_0}\|_{\ell_1} + \|h_{\mathcal{T}_0^\complement}\|_{\ell_1} \\ &\leq 2\|h_{\mathcal{T}_0}\|_{\ell_1} \\ &\leq 2\sqrt{S}\|h_{\mathcal{T}_0}\|_{\ell_2} \\ &\leq \frac{8\sqrt{L}N_{\max}}{\kappa_1^2(S)\sigma_{\min}(A)}\lambda S,\end{aligned}$$

which proves inequality (S14). Furthermore, combining (S17) with (S16) establishes inequality (S15). □

## S1.9 Lemma 8

**Lemma 8.** *Assume all the conditions in Lemma 7. If $\kappa_2(S,S) > 0$ is satisfied, then we have*

$$\|\hat{\gamma} - \gamma^*\|_{\ell_2} \leq \frac{8\sqrt{L}N_{\max}}{\kappa_2^2(S,S)\,\sigma_{\min}(A)}\,\lambda\sqrt{S}, \tag{S18}$$

*and*

$$\|\hat{\eta} - \eta^*\|_{\ell_2} \leq \frac{8\sqrt{L}N_{\max}}{\kappa_2^2(S,S)\,\sigma_{\min}^2(A)}\,\lambda\sqrt{S}. \tag{S19}$$

*Proof of Lemma 8.* From inequality (S16), we have

$$\frac{1}{n}\|Vh\|_{\ell_2}^2 \leq \frac{4\sqrt{L}N_{\max}}{\sigma_{\min}(A)}\lambda\sqrt{S}\|h_{\mathcal{T}_{01}}\|_{\ell_2}.$$

Since $\kappa_2(S,S) > 0$, by Definition 4, the restricted eigenvalue condition implies

$$\frac{1}{n}\|Vh\|_{\ell_2}^2 \geq \kappa_2^2(S,S)\|h_{\mathcal{T}_{01}}\|_{\ell_2}^2.$$

Combining the two inequalities yields

$$\|h_{\mathcal{T}_{01}}\|_{\ell_2} \leq \frac{4\sqrt{L}N_{\max}}{\kappa_2^2(S,S)\sigma_{\min}(A)}\lambda\sqrt{S}.$$

From Lemma 2, specifically inequalities (S3) and (S4), we have

$$\|h_{\mathcal{T}_{01}^\complement}\|_{\ell_2} \leq \frac{1}{\sqrt{S}}\|h_{\mathcal{T}_0^\complement}\|_{\ell_1} \leq \|h_{\mathcal{T}_0}\|_{\ell_2} \leq \|h_{\mathcal{T}_{01}}\|_{\ell_2},$$

and thus,

$$\|h\|_{\ell_2} \leq \|h_{\mathcal{T}_{01}^{\complement}}\|_{\ell_2} + \|h_{\mathcal{T}_{01}}\|_{\ell_2} \leq 2\|h_{\mathcal{T}_{01}}\|_{\ell_2}.$$

Therefore,

$$\|h\|_{\ell_2} \leq \frac{8\sqrt{L}N_{\max}}{\kappa_2^2(S,S)\sigma_{\min}(A)}\lambda\sqrt{S},$$

which proves inequality (S18).

Finally, since $\hat{\eta} - \eta^* = A^+ h$, we have

$$\|\hat{\eta} - \eta^*\|_{\ell_2} = \|A^+ h\|_{\ell_2} \leq \frac{1}{\sigma_{\min}(A)}\|h\|_{\ell_2},$$

which establishes inequality (S19). □

## S1.10 Proof of Theorem 2

*Proof of Theorem 2.* From Lemmas 1 and 7, we know that inequalities (S14) and (S15) hold simultaneously with probability at least $1 - p^{1-C^2/2}$, for some constant $C > \sqrt{2}$ and

$$\lambda = C\sigma\sqrt{\frac{\log p}{n}} + M\mathcal{E}_b.$$

Observe that

$$\|\hat{\eta} - \eta^*\|_{\ell_2} \leq \frac{\|\hat{\gamma} - \gamma^*\|_{\ell_2}}{\sigma_{\min}(A)} \leq \frac{\|\hat{\gamma} - \gamma^*\|_{\ell_1}}{\sigma_{\min}(A)}.$$

Combining this with (S14) and inequality (S13) establishes the bound in (11).

Now consider prediction error. Similar from the proof of Theorem 1, we have

$$\|\hat{\mu} - \mu\|_n \leq \frac{1}{\sqrt{n}}\|X(\hat{\eta} - \eta^*)\|_{\ell_2} + \|v\|_n$$

and

$$\|v\|_n \leq M\mathcal{E}_b.$$

Also, from Lemma 7, we have

$$\frac{1}{n}\|X(\hat{\eta}-\eta^*)\|_{\ell_2}^2 = \frac{1}{n}\|V(\hat{\gamma}-\gamma^*)\|_{\ell_2}^2 \\ \leq \frac{16LN_{\max}^2}{\kappa_1^2(S)\sigma_{\min}^2(A)}\lambda^2 S,$$

which implies

$$\frac{1}{\sqrt{n}}\|X(\hat{\eta}-\eta^*)\|_{\ell_2} \leq \frac{4\sqrt{L}N_{\max}}{\kappa_1(S)\,\sigma_{\min}(A)}\lambda\sqrt{S},$$

and this yields the prediction bound in (12).

Finally, if the stronger restricted eigenvalue condition $\kappa_2(S,S) > 0$ holds, then inequality (13) follows directly from (S13) and Lemma 8. □

# S2 DISCUSSION ON THE RE CONDITIONS

## S2.1 Sufficient Condition for the RE Conditions

The following Lemma 9 states that **(A4)** is a sufficient condition for both **(A5)** and **(A6)**.

**Lemma 9** (cf. Lemma 4.1 in Bickel et al. (2009))**.** *Suppose that* $\phi_{2S} - \theta_{S,2S} > 0$*. Then both* $\kappa_1(S) > 0$ *and* $\kappa_2(S,S) > 0$*.*

*Proof of Lemma 9.* Let $h \in \mathbb{R}^L$ be any vector satisfies that $\|h_{\mathcal{T}_0^\complement}\|_{\ell_1} \leq \|h_{\mathcal{T}_0}\|_{\ell_1}$. Following the construction used in the proof of Lemma 4, we partition $\mathcal{T}_0^\complement$ into disjoint subsets of size at most $S$, i.e.,

$$\mathcal{T}_0^\complement = \bigcup_{k=1}^{K} \mathcal{T}_k,$$

where $|\mathcal{T}_k| = S$ for $k = 1, \ldots, K-1$, and $|\mathcal{T}_K| \leq S$. The sets $\mathcal{T}_k$ are defined recursively: $\mathcal{T}_1$ contains the indices of the $S$ largest (in magnitude) entries of $h$ outside $\mathcal{T}_0$, $\mathcal{T}_2$ the next largest $S$, and so on. Define $\mathcal{T}_{01} = \mathcal{T}_0 \cup \mathcal{T}_1$.

Let $\mathcal{W} \subset \mathbb{R}^n$ be the linear span of $\{V^j : j \in \mathcal{T}_{01}\}$, and let $P_{\mathcal{W}}$ denote the orthogonal projection onto $\mathcal{W}$. From inequality (S8), we have

$$\frac{\|P_{\mathcal{W}} V h_{\mathcal{T}_k}\|_{\ell_2}}{\sqrt{n}} \leq \frac{\theta_{S,2S}}{\sqrt{\phi_{2S}}} \|h_{\mathcal{T}_k}\|_{\ell_2}, \quad \text{for } k \geq 2.$$

Now consider the lower bound:

$$\begin{aligned}
\|P_{\mathcal{W}} V h\|_{\ell_2} &\geq \|P_{\mathcal{W}} V h_{\mathcal{T}_{01}}\|_{\ell_2} - \left\| \sum_{k=2}^{K} P_{\mathcal{W}} V h_{\mathcal{T}_k} \right\|_{\ell_2} \\
&= \|V h_{\mathcal{T}_{01}}\|_{\ell_2} - \sum_{k=2}^{K} \|P_{\mathcal{W}} V h_{\mathcal{T}_k}\|_{\ell_2},
\end{aligned}$$

where we used the fact that $P_{\mathcal{W}} V h_{\mathcal{T}_{01}} = V h_{\mathcal{T}_{01}}$ by construction.

Therefore,

$$\frac{\|P_{\mathcal{W}} V h\|_{\ell_2}}{\sqrt{n}} \geq \sqrt{\phi_{2S}} \|h_{\mathcal{T}_{01}}\|_{\ell_2} - \sum_{k=2}^{K} \frac{\theta_{S,2S}}{\sqrt{\phi_{2S}}} \|h_{\mathcal{T}_k}\|_{\ell_2}.$$

Next, we observe that

$$\sum_{k=2}^{K} \|h_{\mathcal{T}_k}\|_{\ell_2} \leq \frac{1}{\sqrt{S}} \|h_{\mathcal{T}_0^{\complement}}\|_{\ell_1} \leq \frac{1}{\sqrt{S}} \|h_{\mathcal{T}_0}\|_{\ell_1} \leq \|h_{\mathcal{T}_0}\|_{\ell_2} \leq \|h_{\mathcal{T}_{01}}\|_{\ell_2}.$$

Substituting this bound yields

$$\frac{\|P_{\mathcal{W}} V h\|_{\ell_2}}{\sqrt{n}} \geq \sqrt{\phi_{2S}} \left(1 - \frac{\theta_{S,2S}}{\phi_{2S}}\right) \|h_{\mathcal{T}_{01}}\|_{\ell_2}.$$

Finally, since projections cannot increase norm, we have

$$\frac{\|V h\|_{\ell_2}}{\sqrt{n}} \geq \frac{\|P_{\mathcal{W}} V h\|_{\ell_2}}{\sqrt{n}},$$

and therefore,

$$\frac{\|V h\|_{\ell_2}}{\sqrt{n}} \geq \sqrt{\phi_{2S}} \left(1 - \frac{\theta_{S,2S}}{\phi_{2S}}\right) \|h_{\mathcal{T}_{01}}\|_{\ell_2}.$$

This inequality verifies the restricted eigenvalue conditions in Definitions 3 and 4, and hence $\kappa_1(S) > 0$ and $\kappa_2(S, S) > 0$. □

## S2.2 Illustrative Example for the RE Conditions

Suppose we observe a $2 \times 2$ grid over the predictor surface. On this grid, we adopt the piecewise constant basis, which gives $p = 4$ and leads to a basis matrix $\Phi = I_4$. Set the sparsity-smoothness trade-off parameter to $w = 1$, and let the difference orders in both directions be $d_1 = d_2 = 1$. Under this setting, the matrix $A$ becomes

$$A = \begin{bmatrix} 1 & 0 & 0 & 0 \\ 0 & 1 & 0 & 0 \\ 0 & 0 & 1 & 0 \\ 0 & 0 & 0 & 1 \\ 1 & -1 & -1 & 1 \end{bmatrix}.$$

We generate a matrix $X \in \mathbb{R}^{4\times 4}$ with each entry independently drawn from $\mathcal{N}(0, 1)$:

$$X = \begin{bmatrix} 0.7533 & -2.2169 & -1.7922 & 0.8647 \\ 2.0144 & 0.7584 & -0.0420 & -1.7202 \\ -0.3551 & -1.3062 & 2.1500 & 0.1341 \\ 2.0282 & -0.8025 & -1.7702 & -0.0758 \end{bmatrix}.$$

This yields the matrix $V = XA^+$:

$$V = \begin{bmatrix} -0.3721 & -1.0915 & -0.6668 & -0.2608 & 1.1254 \\ 2.0988 & 0.6740 & -0.1265 & -1.6357 & -0.0844 \\ -0.1422 & -1.5192 & 1.9371 & 0.3471 & -0.2130 \\ 1.1231 & 0.1025 & -0.8652 & -0.9808 & 0.9050 \end{bmatrix}.$$

Next, we generate a random vector $h \in \mathbb{R}^5$ such that $\|h_{\mathcal{T}_0^\complement}\|_{\ell_1} \leq \|h_{\mathcal{T}_0}\|_{\ell_1}$, where $|\mathcal{T}_0| = 1$. Let $\mathcal{T}_1$ be the index corresponding to the largest (in magnitude) entry of $h$ outside $\mathcal{T}_0$, and define $\mathcal{T}_{01} = \mathcal{T}_0 \cup \mathcal{T}_1$. We then compute the two ratios:

$$\frac{\|Vh\|_{\ell_2}}{\sqrt{4}\|h_{\mathcal{T}_0}\|_{\ell_2}} \quad \text{and} \quad \frac{\|Vh\|_{\ell_2}}{\sqrt{4}\|h_{\mathcal{T}_{01}}\|_{\ell_2}}.$$

This process is repeated 10,000 times, and the smallest values are recorded. The resulting numerical estimates are $\kappa_1(1) \approx 0.3051$ and $\kappa_2(1,1) \approx 0.2594$.

# S3 DETAILS ON SIMULATION AND REAL DATA STUDY

## S3.1 A Refitting Step

Parameter selection may be performed via information criteria (such as AIC and BIC) when model simplicity is prioritized. Under piecewise constant bases, aligning the estimation grid with basis pieces allows the degrees of freedom to be defined as the size of active set $\mathcal{A}(\hat{\eta}) = \{j : \hat{\eta}_j \neq 0\}$. Such information criteria works well when the sample size is large, but may lead to suboptimal selection in small samples; cross-validation is then preferred.

Large $\lambda$ values, especially when selected by information criteria, can introduce bias. To correct this, one may apply a refitting step, regulating smoothness only over the active set where $\hat{\beta}(t,s) \neq 0$, similar to the Gauss-Dantzig Selector (Candes and Tao (2007)).

Given $\hat{\eta}$ from (6), define

$$\hat{\mathcal{I}}_0 = \{(t,s) \in \mathcal{I} : b^\top(t,s)\hat{\eta} = 0\},$$

and solve

$$\min_{\eta \in \mathbb{R}^p} \|A\eta\|_{\ell_1}, \quad \text{subject to} \quad \left\| \frac{1}{n} N^{-1} X^\top (y - X\eta) \right\|_{\ell_\infty} \leq \lambda \quad \text{and} \quad \Phi_{\hat{\mathcal{I}}_0}^\top \eta = 0.$$

Here, $w = 0$ is set in $A$ to avoid additional shrinkage, and $\lambda$ is re-tuned via cross-validation.

## S3.2 Tuning Details for the Proposed Method in Simulation Study

In the simulation study, the tuning parameters $(\lambda, w)$ were selected jointly using an independent validation set. We considered a 10-point grid for $\lambda$ of the form

$$\lambda = 2^u \sqrt{\frac{2\log(p)}{n}},$$

where $u$ is equally spaced on an interval $[-C_\ell, -C_u]$ with $C_\ell > C_u > 0$, $p$ is the number of basis coefficients ($p = 100$ for $B_1$ and $p = 400$ for $B_2$), and $n$ is the training sample size. The range $[-C_\ell, -C_u]$ was allowed to vary across simulation settings (basis choice and signal type) to ensure that the candidate grid brackets the validation-selected optimum. Specifically, we use $u \in [-12, -8]$ for $\beta_1$ under both $B_1$ and $B_2$; $u \in [-8, -4]$ for $\beta_2$ under basis $B_1$; $u \in [-7, -3]$ for $\beta_3$ under basis $B_1$; $u \in [-6, -2]$ for $\beta_2$ and $\beta_3$ under basis $B_2$.

Our method was empirically insensitive to the sparsity–smoothness weight $w$. Accordingly, we tuned $w$ over the candidate set

$$w = c\, N_{\max}\sqrt{L}, \qquad c \in \{1, 10, 20, 30\},$$

where $L$ is the number of rows of matrix $A$ and $N_{\max}$ is the maximum column $n$-norm of the design matrix $X$.

We found that $d = 2$ and $d = 3$ produced very similar MSEs in this setting when using the basis $B_1$, whereas $d = 3$ typically achieved the smallest MSE when using the basis $B_2$ from preliminary experiments. Accordingly, the differencing order was fixed in the main simulation to (i) reduce the computational burden of introducing an additional tuning dimension and (ii) generate locally quadratic-like surfaces under the chosen basis representations. Specifically, $d_1 = d_2 = 2$ was used for $B_1$ and $d_1 = d_2 = 3$ for $B_2$, where the higher-order differencing helps mitigate blocky artifacts.

To clarify how the differencing order was selected, an additional simulated experiment was conducted in which only $d$ was tuned. The data-generating mechanism matched the main simulation setting: the true coefficient surface was $\beta_1$, the predictor surface was generated from $P_1$, and the sample size, SNR level, spatial grid, and the two basis choices (i.e., $B_1$ (tensor-product cubic B-spline) and $B_2$ (piecewise constant)) were kept the same.

For simplicity, the symmetric choice $d_1 = d_2 = d$ was imposed and the candidate set was restricted to $d \in \{0, 1, 2, 3\}$. To isolate the effect of $d$, the remaining tuning parameters were fixed at

$$\lambda \;=\; 2^{-9}\sqrt{\frac{2\log(p)}{n}}, \qquad w \;=\; \sqrt{L},$$

where $p$ is the number of basis coefficients and $L$ is the dimension of $A\eta$ in (6). For each dataset replicate, $d$ was selected via 5-fold cross-validation by minimizing the mean squared prediction error on the validation folds. The simulation was repeated over 100 independent runs, and the empirical selection frequencies were reported in Table S1.

| Basis | $d = 0$ | $d = 1$ | $d = 2$ | $d = 3$ |
|---|---|---|---|---|
| $B_1$ (tensor-product cubic B-spline) | 1 (1%) | 28 (28%) | 39 (39%) | 32 (32%) |
| $B_2$ (piecewise constant) | 0 (0%) | 2 (2%) | 7 (7%) | 91 (91%) |

Table S1: Selection frequency of differencing order $d$ over 100 repeated simulation runs.

## S3.3 Tuning Details for the Proposed Method in the Real Data Study

In the real-data study, the tuning parameters $(\lambda, w, d)$ were selected jointly via 10-fold cross-validation. The differencing order was restricted to $d \in \{1, 2, 3\}$, see Table S2.

We selected $\lambda$ over a 20-point grid of the form

$$\lambda = 2^{u}\sqrt{\frac{2\log(p)}{n}}, \qquad u \text{ equally spaced on } [-6, -1],$$

where $p$ is the number of basis coefficients ($p = 36$ for $B_1$ and $p = 120$ for $B_2$) and $n = 55$ is the training sample size.

The sparsity-weight parameter was selected over the candidate set $w = c\,N_{\max}\sqrt{L}$, where $c \in \{10, 20, 30\}$. Cross-validation selected $\lambda \approx 0.0503$, $w \approx 2.8091$, $d = 3$ for the final model fit with $B_1$ and $\lambda \approx 0.0582$, $w \approx 5.3785$, $d = 3$ with $B_2$.

| Fold | 1 | 2 | 3 | 4 | 5 | 6 | 7 | 8 | 9 | 10 |
|---|---|---|---|---|---|---|---|---|---|---|
| Selected $d$ with basis $B_1$ | 1 | 1 | 2 | 1 | 1 | 3 | 1 | 2 | 3 | 3 |
| Selected $d$ with basis $B_2$ | 1 | 3 | 3 | 2 | 3 | 3 | 3 | 3 | 3 | 1 |

Table S2: Fold-wise selected differencing order $d$ in the real-data analysis (10-fold CV).

## S3.4 Tuning Details for P-splines and FPCR Methods

The smoothing parameter (denoted by $\rho$) for the P-splines method was selected over grid of the form $\rho = 2^u \rho_0$, where $\rho_0 = n^{-2m/(4m+1)}$ with $m = 2$ and $u$ from a evenly spaced 20-point candidate grid which brackets the validation/CV optimum; we used $u \in [-26, -6]$ in the simulation study and $u \in [0, 10]$ in the real-data study. For FPCR, the number of components was selected to capture $100\,pve\%$ of the variance, where $pve$ was chosen from $\{0.90, 0.91, \ldots, 0.99\}$ for both simulation and real data studies.

For the real data study, the selected $\rho$ for P-splines method is around 0.7249, and the selected $pve$ for FPCR is 0.92,

## S3.5 Additional Figures for the Real Data Study

The coefficient surfaces in Figure 5 are displayed on the same $10^\circ \times 10^\circ$ grid used to construct the predictors: The raw SST is provided on a $2^\circ \times 2^\circ$ grid, but for numerical stability the domain $\Omega$ is partitioned into $10^\circ \times 10^\circ$ cells and the cell-averaged MAM-SST is used as the scalar-on-image predictor. Therefore, the plotted surfaces in the main text are discretized representations at the resolution of the design.

To complement the grid-level coefficient maps in Figure 5, here in Figure S1 we provide a finer-resolution visualization (on the original $2^\circ \times 2^\circ$ grid) of the estimated coefficient surfaces from P-splines and $\mathrm{GDS}_{B_1}$. The MAM-SST field and its association with log(Hurricane Counts) are displayed on the same finer spatial partition, and the coefficient surfaces from P-splines and $\mathrm{GDS}_{B_1}$ are evaluated from their basis representations on the corresponding grid, with contour overlays to highlight local smoothness patterns.

The finer-gridded visualizations are not included for FPCR and $\mathrm{GDS}_{B_2}$ because the fitted coefficients are naturally defined on the $10^\circ \times 10^\circ$ grid for analysis, rather than as an evaluable smooth basis expansion. Specifically, in FPCR the coefficient estimate is returned by the *refund::fpcr* on the discretized grid, so producing a finer-grid contour plot would require an additional interpolation step that is not part of the fitted model and can be sensitive to the interpolation choice. For $\mathrm{GDS}_{B_2}$, the coefficient is estimated on a piecewise-constant (cell-level) basis by construction; evaluating it on a finer grid does not add statistical resolution beyond the $10^\circ \times 10^\circ$ cells, and any contours would again be purely post-processing. This visualization is intended purely for interpretability; the model fitting and prediction remain based on the analysis-grid predictors described in the main text.

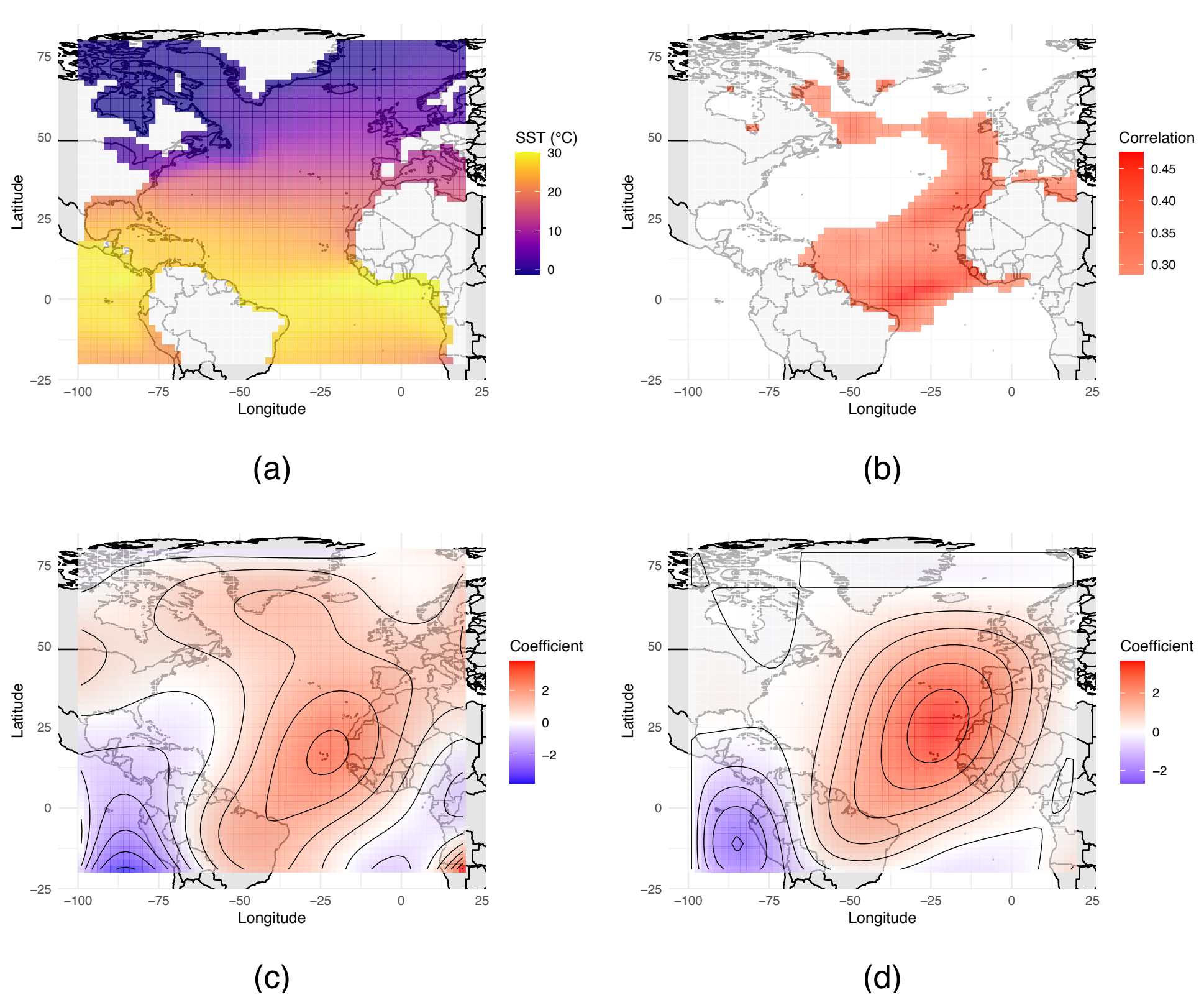

Figure S1: Finer-grid visualizations for the hurricane application. (a) MAM-SST in 2023, computed on the finer spatial partition. (b) Regions with statistically significant correlations between MAM-SST and log(Hurricane Counts), based on Student's $t$-tests with Benjamini–Hochberg FDR control at level 0.05. (c) P-spline coefficient surface evaluated from the fitted tensor-product B-spline representation, with contour overlays. (d) $\text{GDS}_{B_1}$ coefficient surface evaluated from the fitted basis representation, with contour overlays.